\documentstyle[aas2pp4]{article}

\lefthead{Da Costa {\it et al.}}
\righthead{WFPC2 Observations of And I}

\begin{document}

\title{The Dwarf Spheroidal Companions to M31: WFPC2 Observations of 
Andromeda I\footnote{Based on observations with the NASA/ESA {\it 
Hubble Space Telescope}, obtained at the Space Telescope Science Institute,
which is operated by the Association of Universities for Research in 
Astronomy, Inc., (AURA), under NASA Contract NAS 5-26555.}}

\author{G. S. Da Costa}
\affil{Mount Stromlo and Siding Spring Observatories, The Australian National
University, Private Bag, Weston Post Office, ACT 2611, Australia}
\authoremail{gdc@mso.anu.edu.au}

\author{T. E. Armandroff}
\affil{Kitt Peak National Observatory, National Optical Astronomy 
Observatories\footnote{The National Optical Astronomy Observatories
are operated by AURA, Inc., under cooperative agreement with the National
Science Foundation.},\\
P.O. Box 26732, Tucson, Arizona 85726}
\authoremail{armand@noao.edu}

\author{Nelson Caldwell}
\affil{F. L. Whipple Observatory, Smithsonian Institution, P.O. Box 97,
Amado, Arizona 85645}
\authoremail{caldwell@flwo99.sao.arizona.edu}

\and

\author{Patrick Seitzer}
\affil{Department of Astronomy, University of Michigan, Ann Arbor, Michigan
48109}
\authoremail{seitzer@astro.lsa.umich.edu}


\vspace{15.0mm}

\begin{center}
\large{Accepted for Publication in the Astronomical Journal, December 1996 
issue}
\end{center}

\vspace*{\fill}

\pagebreak

\begin{abstract}
Images have been obtained with the {\it Hubble Space Telescope} WFPC2 camera 
of Andromeda I, a dwarf spheroidal (dSph) galaxy that lies in the
outer halo of M31.  The resulting color-magnitude diagrams reveal for the 
first time the morphology of the horizontal branch in this system.  We find
that, in a similar fashion to many of the galactic dSph companions,
the horizontal branch (HB) of And~I is predominantly red.  Combined with the
metal abundance of this dSph, this red HB morphology indicates that And I
can be classified as a ``second parameter'' system in the outer halo of M31.
This result then supports the hypothesis that the outer halo of M31
formed in the same extended chaotic manner as is postulated for the 
outer halo of the Galaxy.  

In addition to the red HB stars, blue HB and RR Lyrae variable stars are also
found in the And I color-magnitude diagram.  The presence of these stars
indicates that And I contains a minority population whose age is
comparable to that of the galactic globular clusters.  We estimate, however,
that the bulk of the stellar population in And~I is $\sim$10 Gyr old.  Thus,
again like many of the galactic dSphs, there is clear evidence for an
extended epoch of star formation in And I\@.  A radial
gradient in the And I HB morphology has also been discovered
in the sense that there are
relatively more blue HB stars beyond the galaxy's core radius.  This may be
evidence for more centrally concentrated star formation after the
initial episode.  Similar HB morphology gradients have also been 
identified in two of three galactic dSphs studied.

The mean magnitude of the blue HB stars suggests that And I lies 
along the line-of-sight  
at the same distance as M31 to within $\sim\pm$70 kpc.
Consequently, the true distance of And I from the
center of M31 is between $\sim$45 and $\sim$85 kpc, with the higher
estimates being more likely.  Such distances are
comparable to the galactocentric distances of the nearer
Milky Way dSph companions Ursa Minor, Draco and
Sculptor.  From the mean color of the lower giant branch, the mean metal
abundance of And I is estimated as [Fe/H] = --1.45
$\pm$ 0.2 dex, while the presence of an internal abundance spread with 
total range of $\sim$0.6 dex is suggested by the intrinsic color width of 
the upper giant branch.  A small population of faint blue stars, which we
identify as blue stragglers, is also present.

\end{abstract}


\section{Introduction}

The nine dwarf spheroidal (dSph) galaxy companions to our Galaxy are, because
of their proximity, the most easily studied examples of what may be the most
common type of galaxy in the Universe.  These systems have been probed in
greater and greater detail in the last decade, and though their evolutionary
history is far from being completely understood, they are all now 
relatively well observed
objects (see recent reviews, e.g. \markcite{GD92}Da~Costa 1992, 
\markcite{RZ93}Zinn 1993a, \markcite{MM96}Mateo 1996).  But within the Local
Group there are at least three other objects that are classified as dwarf
spheroidal galaxies: the dSph companions to M31.  These galaxies, known as
And~I, II and III, lie at projected distances of $\sim$45, 130 and 60 kpc,
respectively, from the center of M31.  {\it Thus in the same way as the 
galactic dSphs are systems in the outer halo of the Galaxy, the And dSphs
are systems in the outer halo of M31}.

Recent work on the And dSphs has established that there is a considerable 
degree of similarity between the M31 dSph system and that of the Galaxy.  For
example, \markcite{CA92}Caldwell {\it et al.\ }(1992) determined accurate 
surface brightness
profiles for the And systems and used these data to show that they follow 
the same relations between absolute magnitude, central surface
brightness and core radius as do the galactic dSphs 
(see also \markcite{AT94}Armandroff 1994).
Ground-based color-magnitude diagrams that cover the upper 2 or 3 magnitudes
of the giant branch are also available for the three And systems: And I
(\markcite{MK90}Mould \& Kristian 1990, hereafter MK90), And II 
(\markcite{KO93}K\"{o}nig {\it et al.\ }1993, see also Armandroff 1994) 
and And~III (\markcite{AD93}Armandroff {\it et al.\ }1993).   These studies
permit estimates of the mean abundance\footnote{Unless otherwise qualified, 
the terms ``abundance'' and ``[Fe/H]'' should be understood as indicating the 
overall abundance of the elements heavier than hydrogen and helium.}, the
distance, and the intermediate-age
population fraction for these galaxies.  All three systems appear to be at,
or close to, the distance of M31 and the mean abundances range from [Fe/H]
$\approx$ --2.0 for And III to [Fe/H] $\approx$ --1.4 for And I\@.  Further, 
the
three dSphs follow the same relation between mean abundance and absolute
magnitude as do the galactic dwarf spheroidals.  As for
intermediate-age population fraction, Armandroff {\it et al.\ }(1993) find
10 $\pm$ 10\% for And I and And III, but for And II this fraction is 
not yet well
established.  However, since intermediate-age upper-AGB carbon
stars have been identified in this system 
(\markcite{AG85}Aaronson {\it et al.\ }1985, see also Armandroff 1994), the
intermediate-age population fraction for this dSph is
undoubtedly larger than that for And I or And III.

Apart from age determinations via measurement of turnoff luminosities, the
major characteristic of the M31 dSphs that is lacking compared to
the galactic dSphs, is information on the morphology of their horizontal
branches.  The galactic dSphs exhibit a wide range of horizontal branch (HB)
morphologies.  For example, the Draco and Ursa Minor systems have similar
abundances and absolute magnitudes yet their HB morphologies are markedly
different: that of Draco being much redder than that of Ursa Minor.  Compared
to the majority of galactic globular clusters of similar abundance to these
systems, it is
Draco that is anomalous.  Draco is said to exhibit the so-called ``second
parameter effect'' in possessing a redder HB than would be expected for its
metal abundance.  This second parameter effect is common among the Galaxy's 
dSph companions and it is also particularly prevalent among the globular
clusters that lie, like the dSphs, in the outer regions of the galactic halo.

Considerable effort has gone into attempting to understand the origin of the
second parameter effect and its dependence on galactocentric distance.  It is
now generally accepted that the phenomenon is a manifestation of age 
differences that occur in the galactic halo (e.g.\ \markcite{LD94}Lee, 
Demarque \& Zinn 1994).  Indeed this age difference interpretation of the
diversity of HB types seen in the outer halo of the Galaxy is a cornerstone
of the chaotic halo formation model advocated initially by 
\markcite{SZ78}Searle \& Zinn (1978), in which the galactic halo is built
out of the destruction of satellite galaxies over an extended interval of
perhaps $\sim$3 to 5 Gyr.  {\it Does this prolonged formation scenario apply
also to the halo of M31?}  An initial attack on this problem can be made by
determining the HB morphologies of the most readily identifiable objects in
the outer halo of M31, the And dSph galaxies.  Since the mean metal abundances
of these systems are well established, we can accurately predict the HB
morphology expected if the bulk of the stellar population in these galaxies
is as old as the inner galactic halo globular clusters.  If the 
observations reveal instead a diversity of HB morphologies, then it will be
an indication that the outer halo of M31 may also have formed in the same
drawn out chaotic manner as the outer regions of the halo of our galaxy.

We begin this task by presenting in this paper the results of a 
{\it Hubble Space Telescope} imaging study whose principal aim 
was to determine the HB morphology of the Andromeda~I dwarf 
spheroidal galaxy.  The
observations, made with the WFPC2 camera, are detailed in the next section
together with a description of the photometric analysis applied to the 
images.  The resulting And I color-magnitude diagrams are discussed in detail 
in
Section 3.  The results, both in the context of the formation and the evolution
of this dSph, and in the context of formation scenarios for
the halo of M31, are discussed in Section 4 and summarized in Section 5.

\section{Observations and Photometry}

Andromeda I was imaged with the 
{\it Hubble Space Telescope} Wide Field Planetary Camera 2 (WFPC2, see
\markcite{H95a}Holtzman {\it et al.\ }1995a for a complete description) on
1994 August 11 and again, at the same spacecraft orientation, on 1994 
August 16.  The first set of observations consisted of three 1800 sec
integrations through the F555W (``Wide-V'') filter and six 1800 sec 
integrations through the F450W (``Wide-B'') filter, while the second set
comprised four 1800 sec F555W and six 1800 sec F450W integrations.  For the
second set of observations (as executed) the HST pointing was set to 
place the center of
And I, determined from the surface brightness study of Caldwell {\it et
al.\ }(1992), on the center of the PC1 chip ({\it i.e.\ }aperture PC1-FIX)\@.
The first set of observations (as executed) however, was offset from this
pointing by a small amount, nominally 20.5 PC pixels in both {\it x} 
and {\it y}, to facilitate investigation of the effects of image undersampling
and small scale flat-fielding variations on the resulting photometry.

The observations were carried out successfully and the raw data frames were
bias and dark subtracted, flat-fielded, etc, via the standard STScI
``pipeline'' process.  The STSDAS task {\it gcombine} was then used to combine,
removing cosmic rays, the individual frames into single ``master'' frames
for each of the four filter and pointing combinations.  Two details 
need to be discussed concerning this
combination process.  First, any pixels labeled as ``bad'' in the accompanying
data quality files were flagged as such in the combined frames by setting
their intensities to a constant value well below that of the ``sky''.  While
this produces frames with obvious cosmetic defects, it is a better
representation of the true data status than generating approximate intensities
for bad pixels from their surrounds.  Second, during the second set of
F555W observations, the spacecraft drifted systematically by a small amount
such that the positions of stars on the last frame differ from those on the
first frame by $\sim$0.5 PC pixels in both {\it x} and {\it y}\@.  Such 
systematic drifts did not occur during any of the other sets of observations;
in these cases the positions of the brighter stars on the individual frames
are constant to better than 0.1 PC pixels in both {\it x} and {\it y}\@.  So,
to compensate for the systematic drift, the first and last F555W PC1 frames in
this group of four were shifted by small amounts, using linear interpolation,
before being combined with the other two frames.  However, since the
equivalent offsets
are a factor of two smaller for the WF frames, and since the WF data is less 
well sampled than that for the PC, no offsets were applied to the WF data
before combining.  Finally the 4 $\times$ 800 $\times$ 800 arrays for each
of the 4 master frames were split into separate 800 $\times$ 800 arrays and
then multiplied by the appropriate geometric correction distortion frames as
supplied by STScI\@.  The vignetted regions at the edges of the frames
(0 $\le$ {\it x, y} $\le$ 100 for the PC, 0 $\le$ {\it x, y} $\le$ 75 for the 
WF frames) were then removed.  A mosaic of the combined F555W image from the
second set of observations is shown in Fig.\ 1; note the completely resolved
and relatively uncrowded appearance of this dwarf galaxy, reminiscent of 
ground-based images of galactic dSph systems.  We note also that there are no 
globular star clusters evident in this And I image.  None have been found
on ground-based images either (MK90, Caldwell {\it et al.\ }1992), and so it
seems likely that And I lacks such clusters.  This result is not unexpected
given that And I's luminosity is $\sim$2 magnitudes fainter than those of 
Fornax and Sagittarius, the two galactic dSphs that possess their own globular
cluster systems.

\begin{figure}[htb]
\figurenum{1}
\epsscale{1.00}
\caption{A mosaic of the And I WFPC2 field made from the
combination of 4 1800 sec F555W images.  North is indicated by the direction 
of the arrow and East by the line.  Both indicators are 10$\arcsec$ in
length.}
\end{figure}

Photometry was then carried out on each of the sixteen combined data frames
(2~positions, 2~filters, 4~CCDs) in the same manner.  First, the IRAF/DAOPHOT
routine {\it daofind} was used to find images on the frames.  It was found
that provided the threshold was kept reasonably high ($\sim$5$\sigma$,
where $\sigma$ is the standard deviation of the 
background), the number of spurious objects found was minimized.  Aperture
photometry was then carried out using the {\it daofind} coordinate list as 
input.  For the PC1 frames the measurement aperture was 2.5 pixels in radius
and the annulus in which the sky was determined had inner and outer radii of
6 and 16 pixels, respectively.  For the WF frames the measurement aperture was
2.0 pixels in radius and the sky annulus lay between radii of 5 and 15 pixels.
These measurement apertures are small enough to maximize the signal-to-noise
ratio for faint stars but are not so small that centering uncertainty or
variations in the PSF within the aperture across the frame are significant.  
No redetermination of the image centers was performed and, in all cases, the 
sky value was determined from the mode of the
pixel intensity distribution within the sky annulus.

The first step in the subsequent processing was to compare the aperture
magnitudes for the brighter stars on the two F555W frames and on the two F450W
frames for each CCD\@.  For the F450W frames the agreement is excellent: the
mean values of the differences between the two measures being (0.004, 0.005,
0.001 and -0.001) with standard deviations of (0.034, 0.050, 0.038, 0.039)
for (49, 119, 76 and 110) brighter stars on the (PC1, WF2, WF3, WF4) frames, 
respectively.  As a result, the two sets of magnitudes were averaged for the
stars in common, with the stars detected on only one frame or the other not
retained.  This process removes any remaining spurious images, such as 
``hot'' (high dark current) pixels that do not follow the coordinate offsets
between the real stars.  For the two sets of F555W magnitudes, the 
corresponding magnitude differences (in the sense 
{\it set$_{1}$ -- set$_{2}$}),
standard deviations and sample sizes are (-0.039, -0.034, -0.040, -0.031),
(0.025, 0.036, 0.035, 0.036) and (35, 96, 63, 90), respectively, for the
(PC1, WF2, WF3, WF4) frames.  There is clearly a systematic offset in the
magnitudes here which
we attribute to signal loss occurring during the combination process for the 
individual frames from the second F555W data set, the set during which the
spacecraft drifted.  Consequently, we have applied these mean offsets to the
{\it set$_{2}$} magnitudes before combining them with the magnitudes from
{\it set$_{1}$}\@.  Again only stars measured on both frames were retained.

The next step requires the determination of the aperture corrections to convert
the small measurement aperture magnitudes to magnitudes on 
the WFPC2 system (Holtzman {\it et al.\ }1995b, hereafter H95b), which is
based on standard apertures of 0.5$\arcsec$ radius.  These aperture corrections
are no doubt position dependant within each of the 4 CCD fields, but we lack 
the numerous bright high S/N stars needed to define well this spatial 
variation.
We have therefore proceeded as follows.  For the PC we first calculated 
aperture corrections for the dozen or so bright uncrowded stars 
for which such a determination was reasonably precise.  Then we used 
the {\it Tiny Tim} PSF simulation package (Version 4.0b; 
\markcite{KR93}Krist \& Hasan 1993, \markcite{KR94}Krist 1994) to calculate 
synthetic PSFs
at the locations of these stars.  These synthetic data frames were then 
measured in exactly the same way as the real stars and the resulting aperture
corrections compared with the determinations from the real data.  The 
agreement was quite satisfactory (average mean differences less than 0.01 mag)
so we then used the {\it Tiny Tim} package to calculate, for each filter, 
a 64 $\times$ 64 grid of synthetic PSFs.  This grid was then converted into a 
grid of aperture corrections and the small aperture photometry then corrected 
to the standard aperture size by interpolating in this grid.

For the WF CCDs however, this process was not as satisfactory.  The {\it Tiny
Tim} PSFs gave aperture corrections that were typically 0.05 mag
larger than those calculated from the stars themselves,
indicating that the real data is more centrally condensed than the model PSFs.
We therefore decided to adopt a single aperture correction for each 
WF filter and CCD combination, thereby averaging over any spatial variation.
These mean aperture corrections were determined from typically ten bright
uncrowded stars on each frame with, at least for F450W, the values from the
two datasets being averaged (the sets of stars used on each frame were
generally not identical).  Based on the {\it Tiny Tim} simulations, we expect
that, for approximately the central two-thirds of the frame area, the adopted 
mean aperture corrections will be satisfactory.  Outside this region, and
particularly in the corners, the adopted aperture corrections will be
underestimates (i.e. the true magnitude will be brighter) by up to perhaps
a few hundredths of a magnitude at most.  We note however, that this failure
to account fully for the spatial variation in the aperture corrections on
the WF frames will have only a second order effect on the colors of the stars
since, based on the {\it Tiny Tim} modelling, the spatial variation is not a
strong function of wavelength.

After application of the aperture corrections we can then employ the 
zeropoints 
and gain factors of H95b to place the photometry on the WFPC2 system.  One
further correction needs to be applied however, and that is the correction
for charge transfer effects (H95b).  For the WF frames, we used the correction
recommended in H95b, a simple linear ramp correction amounting to an 
increase in brightness of 0.04 mag for the stars at the highest {\it y} values.
Note again that since this ramp was applied to both the F450W and F555W
magnitudes, the colors of the stars are unaffected.  For the PC however,
plots of color-magnitude diagrams for different {\it y}--sections of the 
frame, after the application of a 0.04 mag/800 pixels ramp to both the
F450W and F555W magnitudes, showed a 
distinct redward shift in the principal sequences with
increasing {\it y} coordinate.  We interpreted this result as indicating that
a 0.04 mag/800 pixels ramp was inadequate for the F450W PC data, which has
a background of only about 7e$^{-}$ on the individual PC1 frames.  A ramp of 
0.10 mag/800 pixels was necessary to remove this effect.
At this stage we now have fully corrected F450W and F555W magnitudes; all that
remains is to combine the lists to produce F450W--F555W colors.  As was the
case for the magnitudes, only those stars that are found in both lists were 
retained, so that in effect, a star has to have been measured on all four
frames (2 filters, 2 pointings) to be included on the final lists.  

Since
we are primarily interested in the morphology of the And I color-magnitude
diagram, we can afford to edit the photometry lists to remove stars that
may have larger than average errors.  The principal source of additional 
error will be that which arises from image crowding.  
Although these frames are by no means crowded,
there are still clearly instances where the light in the measurement aperture
for one star is contaminated by light from a near neighbor.  Accordingly, we
removed from the photometry lists any star whose center lay within 5 (WF)
or 9 (PC) pixels of the center of its nearest neighbor.  These radii, which
reduce the lists by about 15\%, were chosen as the best compromise between
maximizing the number of stars retained and reducing the scatter in the 
color-magnitude diagrams.  Stars that lay near overexposed bright stars 
or on the
mottled background of the few extended background galaxies were also removed
at this stage.  Finally, the stars remaining were visually inspected on the
{\it set$_{2}$} F555W combined frame.  This led to the removal of a few
additional ``stars'', most of which were reclassified as marginally resolved
background galaxies.

The final color-magnitude diagrams (hereafter cmds), containing 
(256, 1074, 952 and 1226)
stars brighter than F555W = 26.7, are shown in the panels of 
Fig.\ 2 for the (PC1, WF2, WF3 and WF4) data, respectively.  There are clearly
no obvious systematic differences between these 4 cmds.  Moreover, calculations
of mean colors and magnitudes at various places in the cmds indicate that
any relative zeropoint differences between the four sets of data 
are at the $\pm$0.01 mag level at worst.  
We have therefore combined the data from the individual CCDs into a single
color-magnitude diagram for Andromeda I containing 3508 stars.  This diagram 
is 
shown in Fig.\ 3\@.  Fig.\ 4 shows the equivalent V, B--V cmd which
results from applying the transformation equations given in H95b to the data
of Fig.\ 3.

\begin{figure}
\figurenum{2}
\epsscale{0.8}
\plotone{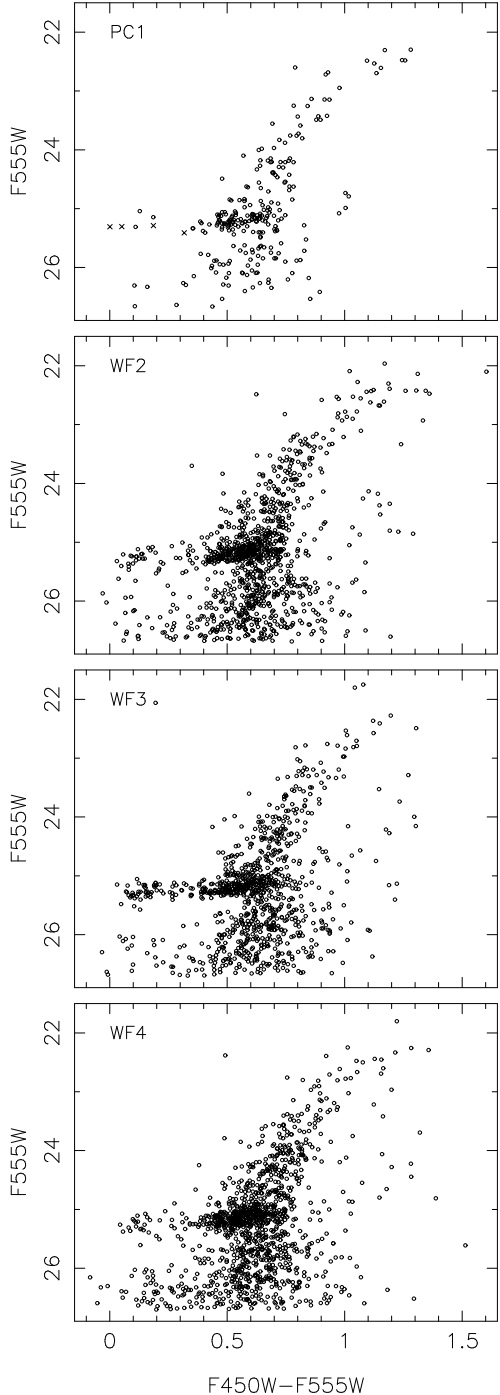}
\caption{The And I color-magnitude diagrams, in the HST 
instrumental system, derived from the data for each WFPC2 CCD\@.  The 
panels are labeled with
the CCD identification.  In the panel for the PC1 chip, the four RR Lyrae
variables found in the PC1 data are plotted as $\times$--symbols.}
\end{figure}

Before discussing the astrophysical results that flow from Fig.\ 3, we turn
first to a brief discussion of the {\it photometric} errors in these 
data.  These
errors are those that arise from the small aperture measurement process, as
distinct from the systematic errors that arise from the uncertainty in the
aperture corrections, in the zeropoint calibration, in the ramp correction,
and so on.  The photometric
errors can be readily investigated by making use of the two sets of data we
have available.  For the stars in Fig.\ 3 then, we returned to the
photometry lists for the two separate pointings and compared the magnitudes 
and colors.  The results of this comparison are given in Table 1 where we 
have tabulated
the average error in the mean of two measures (either magnitude or color) for
the listed magnitude bins.  Three points are worth making in
connection with this table.  First, at the fainter magnitudes (F555W, F450W
$\gtrsim$ 26.0) where the errors are relatively large, the mean errors 
in the magnitudes found
from comparing the independent measures closely approximate the errors expected
on the basis of the photon statistics alone.  In other words, for uncrowded
faint stars, the WFPC2 system is producing photon statistics limited data as
is usually the case with ground-based CCD cameras.  However, at the brighter
magnitudes, this does not seem to be the case.  For the brighter stars the
photon statistics errors are below 0.01 mag yet on the basis of the 
frame-to-frame comparison, the decrease in the errors with increasing
 brightness
flattens out and does not go below $\sim$0.02 mag.  This limit is 
probably the result of a number of inherent effects in the flat-fielding,
dark subtraction and individual frame combination processes.  Fortunately,
the existence of such a limit does not compromise our data in any way.  
Finally, we note that for the majority of stars, $\sigma^{2}_{F450W-F555W}
\lesssim \sigma^{2}_{F450W} + \sigma^{2}_{F555W}$ as might be expected when
the errors are not determined simply by photon statistics.

\begin{figure}[htb]
\figurenum{}
\epsscale{1.00}
\plotone{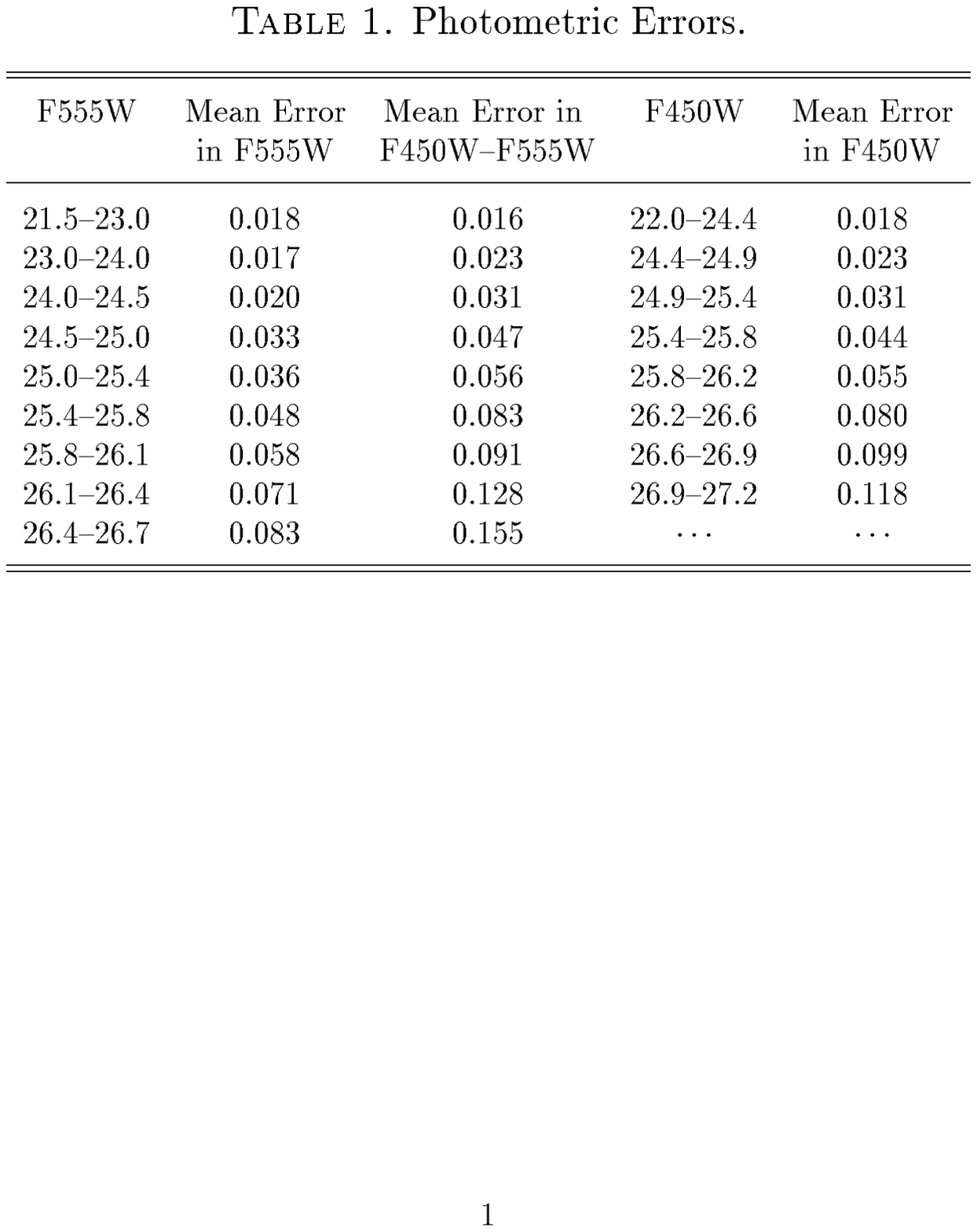}
\end{figure}

\section{Results}

In this section we will discuss the results that can be inferred from the
And I cmd shown in Fig.\ 3, and its equivalent in the standard (V, B--V)
system (Fig.\ 4)\@.  In this process we will first emphasize those results,
such as the morphology of the horizontal branch, that do not depend on the 
photometric calibration of the cmd.  Then we will move to those, 
such as the distance
of And I, that depend primarily on the calibration of the F555W photometry
and its transformation to V magnitudes.  Finally, we will discuss quantities,
such as the mean abundance of And I, which depend on the calibration of both
the F450W and F555W magnitudes as well as on their transformation to
standard B and V magnitudes.  As we shall see, the F450W to B transformation
is less well established than that of F555W to V\@.

\subsection{Color-Magnitude Diagrams of And I}

The morphology of the color-magnitude diagram shown in Fig.\ 3 is instantly
recognizable as that of a basically old stellar population.   A red
giant branch that terminates at F555W $\approx$ 22.3 and F450W--F555W
$\approx$ 1.3 is visible, and there is a dominant red horizontal branch
at V $\approx$ 25.2.  A less well populated blue horizontal branch is also
present.  Stars evolving away from the red HB towards the asymptotic
giant branch are evident to the blue of the giant branch for
V $\gtrsim$ 24.  The three stars well to the blue of the giant branch with
F555W $\approx$ 22.0, F450W--F555W $\approx$ 0.2 through F555W $\approx$ 22.5,
F450W--F555W $\approx$ 0.6 may be post-AGB stars.  At all magnitudes a small 
number of stars are
visible well to the red of the giant branch.  In addition,
there is a small population
of faint (F555W $\gtrsim$ 26.0), blue (F450W--F555W $\lesssim$ 0.25) stars.

\begin{figure}[htb]
\figurenum{3}
\epsscale{1.00}
\plotone{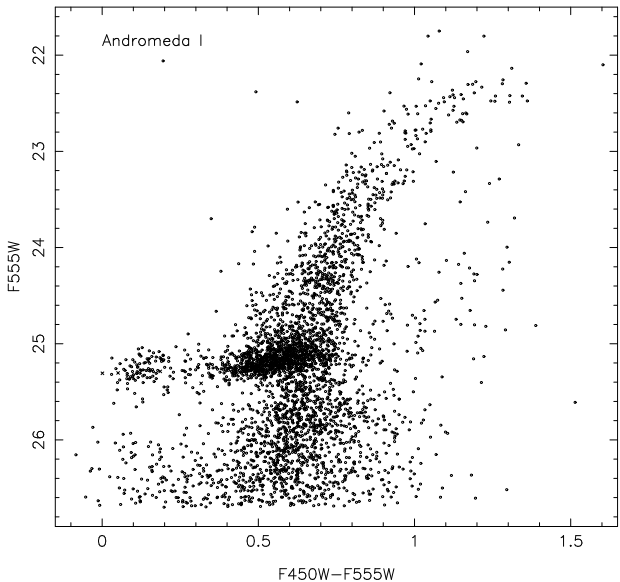}
\caption{The combined Andromeda I color-magnitude 
diagram in the HST instrumental system.  The dominant red horizontal branch
morphology for this dSph galaxy is particularly striking.  The 4 RR Lyrae
variables found in the PC1 data are again shown as $\times$--symbols.}
\end{figure}

Before discussing these And I populations in more detail however, it is
first necessary to consider the likely number of contaminating objects in
Fig.\ 3 (and Fig.\ 4)\@.  Such objects can come from one of three components: 
foreground
galactic stars, stars in the halo of M31, and unresolved background galaxies.
Considering galactic foreground stars first, the model predictions of
\markcite{RB85}Ratnatunga \& Bahcall (1985) suggest $\sim$14 
galactic foreground stars should be found in Fig.\ 4 with B--V $\ge$ 1.3 
and 23 $\le$ V $\le$ 25.  In Fig.\ 4, counting only the stars that are
considerably redder and fainter than the 47 Tuc giant branch (see Fig.\ 8),
there are $\sim$30
to 35 stars in this color and magnitude interval.  Given the uncertainties
in the Ratnatunga \& Bahcall model, and the fact that some of these ``stars''
may be unresolved galaxies, the agreement between the observed and predicted 
numbers is acceptable.  Consequently, we can use the model prediction that
bluer stars are much rarer at these magnitudes to
conclude that there are few, if any, galactic foreground stars with B--V
$\lesssim$ 1.3 in Fig.\ 4 at any magnitude.  Thus the And I giant and
horizontal branches are unaffected by foreground contamination.

To estimate the contribution of M31 halo stars to Figs.\ 3 and 4, we have
proceeded as follows.  And I lies some 3.3$\arcdeg$ from the center of M31
at a position angle of $\sim$135$\arcdeg$ relative to the M31 major axis.  
The data
of \markcite{PV94}Pritchet \& van den Bergh (1994) then suggest that the
halo of M31 has a surface brightness at this location of perhaps 29.5 V
mag/arcsec$^{2}$, though the uncertainty in this number is large 
($\sim\pm$1 mag).  On the other hand, using the surface brightness 
data of Caldwell {\it et al.\ }(1992), 
the average And I surface brightness over the region
contained on the WFPC2 images is approximately 25.4 V mag/arcsec$^{2}$.  Then,
under the reasonable assumption of similar stellar populations, the
ratio of these two surface brightnesses predicts that And I stars should 
outnumber M31 halo stars by about 45 to 1, indicating a minor degree
of contamination.  We can attempt to verify this estimate by making use of
the results of \markcite{DH94}Durrell {\it et al.\ }(1994, see also 
Mould \& Kristian 1986).  These authors indicate that the
mean abundance of the M31 halo is similar to that of the galactic globular
cluster 47 Tuc, though there is
a significant spread to higher and lower abundances.  In Fig.\ 4 there are
approximately 30 stars with 23 $\le$ V $\le$ 25 that scatter about, and to the
red of, the 47 Tuc giant branch (see Fig.\ 8) and which could reasonably 
be ascribed to the M31 halo,
rather than to And I\@.  After allowing for the M31 halo abundance spread
(Durrell {\it et al.\ }1994), this
suggests that perhaps there are a total of $\sim$40 M31 halo stars with
23 $\le$ V $\le$ 25 in Fig.\ 4\@.  The surface brightness scaling on 
the other hand would
predict $\sim$20 M31 halo stars in the same magnitude interval.  Thus,
as found for the galactic foreground stars, the
``observed'' number is higher than the predicted number, but we are not
seriously concerned with this discrepancy.  Even if the higher figure is valid,
the comparative lack of metal-poor M31 halo giants (Durrell {\it et al.\ }1994)
ensures that the degree of contamination of the And I giant branch 
by (metal-poor) M31 halo stars is too small to affect any of the results.  

\begin{figure}[htb]
\figurenum{4}
\epsscale{1.00}
\plotone{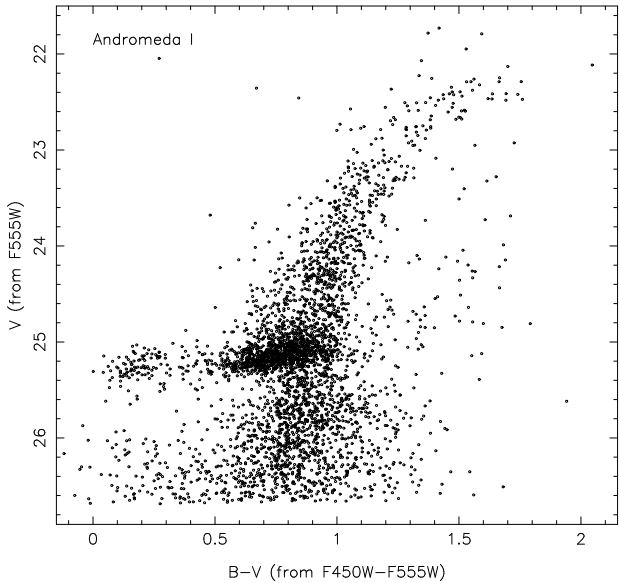}
\caption{The combined Andromeda I color-magnitude 
diagram in the standard (V, B--V) system.  The transformations applied to 
the HST instrumental system photometry are those of 
Holtzman {\it et al.\ }(1995b).}
\end{figure}

As regards the numbers of unresolved background galaxies, the uncertainties
in the actual contributions of the galactic foreground and M31 halo stars
make it difficult to place any meaningful constraints.  We note only that
the results of \markcite{CR95}Casertano {\it et al.\ }(1995), for example,
indicate that E galaxies become smaller and redder with increasing
apparent magnitude.  Thus while any unresolved background galaxies in our 
data may contribute to the ``stars'' that lie to the right of the And I
giant branch in Fig.\ 4, it is unlikely that they will be a major contaminant
of the And I giant and horizontal branches.

\subsubsection{The Horizontal Branch Morphology}

It is immediately apparent from Figs.\ 3 and 4 that the morphology of the
horizontal branch in And I is dominated by red HB stars.  Indeed, the
red end of the HB is not clearly distinguishable from the blue edge of
the red giant branch, as is usually the case in color-magnitude diagrams for 
old
stellar populations.  We believe this effect results from the combination of
a number of factors.  First, as indicated in Table 1, the photometric errors 
in the colors at the magnitude of the HB are approximately $\pm$0.06 mag. 
Thus an intrinsically narrow color width on the red HB would be broadened to
an observed F450W--F555W color range of perhaps 0.24 mag.  
Second, the relatively small 
difference in effective wavelength between the F450W and F555W filters means
that for a given effective temperature difference between a red HB star and
a red giant at the same magnitude, the color difference F450W--F555W will be
smaller than it would be in, for example, the standard B--V system.  Third, as
we will argue below, there is an intrinsic color width to the And I red giant
branch which is probably the result of an internal abundance range.  
Consequently, metal-richer red HB stars will lie closer in color to the
metal-poorer red giants than would be the case for a single abundance
population.  An age range could also play a role here since the HB is redder 
and
the giant branch somewhat bluer in a younger population (e.g.\  
\markcite{SL95}Sarajedini {\it et al.\ }1995)\@.  Consequently, in 
conjunction with any real effective temperature spread among
the And I red HB stars, these factors will combine to blur the distinction
between red HB and red giant branch stars at the level of the horizontal 
branch.  Nevertheless, it is apparent from Figs.\ 2 and 3 that the 
number of stars at the magnitude of the horizontal branch
begins to decrease for colors redder than F450W--F555W $\approx$ 0.6 mag.  We
will therefore take this color as a reasonable estimate of the color blueward
of which core helium burning stars dominate.

We can then quantify the HB morphology of And I as follows.  Define {\it b} as
the number of stars with 25.0 $\le$ F555W $\le$ 25.5 and 0.0 $\le$ F450W--F555W
$\le$ 0.25 and {\it r} as the number of stars in the same magnitude interval
but with 0.35 $\le$ F450W--F555W $\le$ 0.60 mag.  We can then calculate a
HB morphology index {\it i} = {\it b}/({\it b}+{\it r})\@.  This 
index\footnote{We cannot
use the now common HB morphology index (B--R)/(B+V+R), where B, V and R are the
numbers of blue, variable and red HB stars, respectively, because we have 
not yet identified all the variables in the And I cmd shown in Fig.\ 3.} 
({\it cf.\ }\markcite{MI72}Mironov 1973) ranges from zero for a 
pure red HB to unity for
a pure blue HB\@.  From the data of Fig.\ 3, this index has a value {\it i} =
0.13 $\pm$ 0.01 for And I, where the error has been calculated assuming that
both {\it b} and {\it r} are subject to Poissonian statistics.

How then are we to interpret this index?  As we shall see in Sect.\ 
3.3, the mean abundance of And I is [Fe/H] = --1.45 $\pm$ 0.2 dex.  At this
abundance, the cmds of galactic globular clusters which follow the
(HB morphology, [Fe/H]) relation defined by the majority of such clusters,
have relatively many more blue HB stars and many fewer red HB stars than does
And I\@.  For example, the galactic globular cluster M5, which lies on the
(HB morphology, [Fe/H]) relation for the inner-halo (the majority of clusters)
in Lee {\it et al.\ } (1994), and for which [Fe/H] =
--1.40 (\markcite{TA89}Armandroff 1989), has {\it i} $\approx$ 0.74 using
the cmd of \markcite{BI81}Buonanno {\it et al.\ }(1981).  In this sense we
can then say that {\it And I shows the second parameter effect} in the same way
as do many of the galactic dSph systems.  Indeed our data for And I, as plotted
in Fig.\ 4, bears many similarities to the ground-based cmd of 
\markcite{DI93}Demers \&
Irwin (1993) for the galactic dSph companion Leo II\@.  This dSph has a
somewhat lower
mean abundance than And I ([Fe/H] = --1.9, Demers \& Irwin 1993; --1.9 $\pm$
0.2, 
\markcite{SA86}Suntzeff {\it et al.\ }1986; -1.60 $\pm$ 0.25, 
\markcite{MR96}Mighell \& Rich 1996)
but it also shows a predominately red HB\@.  Demers \& Irwin give 
(B--R)/(B+V+R)
= --0.68 for Leo II from which we infer {\it i} $\approx$ 0.15, a value
similar to that
found here for And I\@.  Demers \& Irwin (1993) also note that there is no
obvious break between the red edge of the HB and the red giant branch in
their ground-based cmd.  However, in the Leo II cmds of Mighell \& Rich (1996),
which are derived from WFPC2 F555W and F814W observations and which are
considerably more precise than the Demers \& Irwin (1993) ground-based
data, the red edge of
the HB is separated by $\sim$0.1 in V--I from the red giant branch.

The horizontal branch in Figs.\ 3 and 4 does not, however, consist solely of
red horizontal branch stars.  Instead it extends to quite blue colors and this
suggests that a search for RR Lyrae variables among the bluer HB stars
could prove fruitful.  As noted in Sect.\ 2, the F450W observations consist
of two sets of six 1800 sec exposures separated by an interval of approximately
5.3 days.  Within each set, the exposures are spaced by the HST orbital
period of $\sim$96 min so that the total interval covered by each F450W set of
observations is
$\sim$0.4 day.  These intervals, while not ideal, are adequate to search for RR
Lyrae variables, since such stars have periods of $\lesssim$0.7 day.  
Consequently,
in order to carry out an initial reconnaissance for variable stars 
on the horizontal branch in And I, we have investigated the
variability characteristics of the 14 stars
on the PC frame with 25.0 $\le$ F555W $\le$ 25.5 and 
F450W--F555W $\le$ 0.43 (B--V~$\approx$~0.6).

A number of points have to be taken into account when assessing the results of
this process.  First, the individual data frames are, of course, not free of
cosmic-ray contamination.  As a result, although the aperture photometry was
carried out in the same fashion as for the combined frame (using the centers
determined from the combined frame), each variable candidate had to be 
inspected
visually on each frame to be certain that the aperture measurement was not
affected by cosmic rays.  Typically two of the six possible measurements
from each set of F450W frames had to be discarded.  Second, the individual
photometric errors are in the range (1$\sigma$) $\sim$0.10 - 0.15 mag.
Therefore, when considering the set of (at most twelve) individual 
F450W magnitudes,
variations had to exceed 0.4~-~0.5 mag before they were taken as an
indication of possible intrinsic variability.  Hence we are not likely to
detect any variables whose amplitudes are less than these values.  For the same
reason
we are likely to miss any variable whose maximum occurred outside the two 
$\sim$0.4 day observing windows.  However, despite
these complications, it was quite obvious that four of the 14 candidates varied
with F450W (full) amplitudes ranging from $\sim$0.6 mag to $\sim$1.2 mag.
These stars are identified in the PC1 c-m diagram of Fig.\ 2 by the
use of a $\times$--symbol.  The colors of these variables, from the final
combined frame photometry, are all significantly bluer than the color 
cutoff used to define
the candidates, so that, subject to the caveats given above, it is unlikely 
that we have missed many additional variables.  

For these four stars we have used the individual magnitudes and the 
mid-exposure times as input into the Phase Dispersion Minimization (PDM)
routine within IRAF, in order to generate estimates of their periods.  The 
five day gap between the two sets of observations necessarily
introduces some ambiguity into this process and as a result, we have
resorted to use of the period-amplitude relations exhibited by RR Lyrae
stars in the galactic dSph systems 
(e.g.\ \markcite{KK95}Kaluzny {\it et al.\ }1995 and
references therein) to aid in identifying the appropriate number of cycles
between the two datasets.  Photometry of these stars from the F555W 
observations (3$\times$1800 sec and 4$\times$1800 sec preceding the
corresponding F450W observations) was also employed to constrain the range
of possible periods.  

The results of this process are shown in Fig.\ 5 where we present light curves
for the four variables.  Two of the stars are evidently type ab RR Lyrae
stars while the other two are type c variables.  The periods given for these
stars are accurate only to a few digits in the third significant figure and the
phase is relative to the mid-exposure time of the first 1800 sec F450W 
exposure.  With such a small sample it is impossible to draw any firm
conclusions from the properties of these And I RR Lyrae variables.  
Nevertheless, we note that the
relative number of RRc and RRab stars discovered here in the PC data is not 
incompatible with that seen in
the Sculptor dSph: Kaluzny {\it et al.\ }(1995) have found 89 type~c and 134
type~ab variables in this system.  We also note that, given the relative 
areas of the PC and WF frames, a total population of perhaps 50 to 60 And I
variables should be present in the entire dataset.  The characteristics of 
these stars will be the subject of a subsequent paper.  

\begin{figure}
\figurenum{5}
\epsscale{0.80}
\plotone{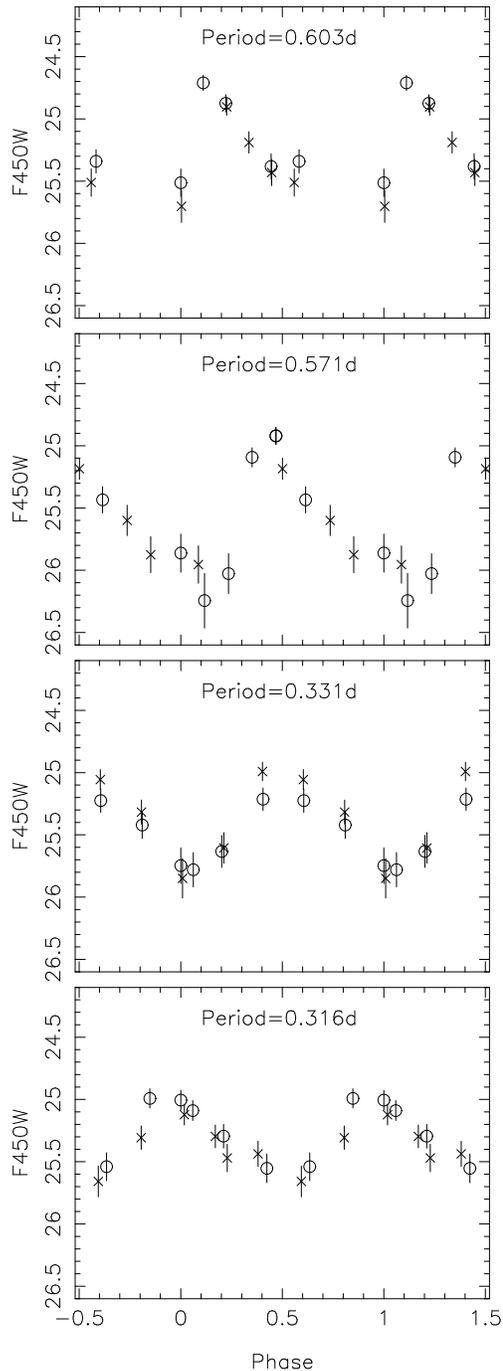}
\caption{F450W light curves for the 4 RR Lyrae 
variables found in the PC1 photometry.  Two cycles are plotted for each 
variable with open circles and 
$\times$--symbols representing magnitudes from the two sets of F450W 
observations.  The error bars are those from the photon statistics.  The
adopted periods are given in each panel.}
\end{figure}

\subsubsection{Radial Gradients?}

Color-magnitude diagram studies of the galactic dSph companions are generally
limited, by the large apparent size of these galaxies, to regions well
within the core radius.  This restriction necessarily constrains any search for
radial gradients whose presence, or confirmed absence, would provide 
important clues to the formation of dSph systems.  For And I, where the 
core radius\footnote{Throughout this paper the term ``core radius'' is taken
to mean the radius at which the projected surface brightness profile reaches
one-half of its central value.  As a result of the low central concentrations
of dSph galaxies, the core radius defined in this way is generally somewhat
smaller than the ``core radius'' parameter derived from a 
\markcite{KI66}King (1966) model
fit to the surface brightness profile data.} as determined from the 
surface photometry study of Caldwell {\it et al.\ }(1992), is 
95~$\pm$~5$\arcsec$, there is sufficient radial coverage in the WFPC2 data
to look somewhat beyond the galaxy's core.  We have therefore sought to 
identify any radial trends present in the characteristics of the cmds 
shown in Figs. 3 and 4.  As a first step, it is necessary to transform 
the ``local''
{\it x} and {\it y} coordinates from the photometry for each WFPC2 CCD to a
single global system.  At the level of precision required here, it is not
necessary to allow for the slight rotations and misalignment of the CCDs nor 
for the minor
scale changes induced by the camera distortions.  Instead a simple global
system was adopted which has {\it x} and {\it y} axes parallel to those of
the PC1 data and which has its origin at the camera apex.  The original
coordinates from the WF CCD photometry were then flipped and changed in
sign as appropriate and those from the PC1 photometry multiplied by the
relative PC to WF scale factor of 0.457 (H95a).  Then, since the center of
And I (also determined from the surface photometry of 
Caldwell {\it et al.\ }1992) was centered on the center of the PC1 field,
it is straightforward to generate cmds 
similar to those of Figs. 3 and 4 for different distances from the galaxy's
center.  Caldwell {\it et al.\ }(1992) have also shown that And I is circularly
symmetric so no allowance for ellipticity is necessary.

We have first investigated the radial dependence of the mean color of the 
giant branch at a number of different luminosities.  This will provide
limits on any radial variation in the mean abundance of And I's dominant 
stellar population.  No evidence for any change with distance from the 
galaxy's center was found.  For example, if we consider just two samples,
namely the stars inside and outside a radius of 900 pixels 
(i.e.\ $\sim$90$\arcsec$ or approximately 1 core radius), then the mean
B--V colors for stars inside the dividing radius are 0.85 $\pm$ 0.01 (std
error of the mean) for 25.5 $\le$ V $\le$ 25.7 and 1.15 $\pm$ 0.02 for 
stars with 23.1 $\le$ V $\le$ 23.3.  For the stars outside the dividing
radius, the corresponding mean colors for the same magnitude intervals are
0.87 $\pm$ 0.02 and 1.16 $\pm$ 0.03, respectively.  This lack of any
significant difference in mean color is found regardless of the radial
binning chosen or of the luminosity range considered.  Anticipating 
the abundance
calibration discussed in Sect.\ 3.3 below, these results imply an upper
limit of $\sim$0.2 dex for any change in the And~I {\it mean}
abundance over a radial
distance from the center to $\sim$1.2 core radii.  This result is consistent 
with those of Caldwell {\it et al.\ }(1992) who found that And I's
integrated B--V color is constant (at the $\pm$1$\sigma$ = 0.02 mag level)
inside a radius of $\sim$120$\arcsec$ or $\sim$1.3 core radii.  We emphasize
though that this result applies to the dominant stellar population in And I;
an abundance gradient could still occur in any minor stellar
population component present.

The second quantity we have investigated for any radial dependence is the
horizontal branch morphology index {\it i} = {\it b}/({\it b}+{\it r})\@.
Unlike
the mean giant branch color where any radial change reflects changes in the
properties of the dominant stellar population, the HB morphology index
is sensitive to changes in the relative numbers of blue HB stars, which are 
only a {\it minor}
component (recall {\it i} = 0.13 $\pm$ 0.01 for the complete sample). 
We first split the total sample radially into four subsamples each containing
roughly equal numbers of stars and then computed the HB morphology indices.  
Intriguingly, the value of {\it i} for the outermost radial bin was
substantially larger (i.e.\ relatively more blue HB stars) than for the other
more centrally located samples.  This led us to conduct a number of 
experiments with
different radial selections.  From these we have ascertained that within
approximately one core radius\footnote{Since the core radius was determined 
from
surface photometry, it corresponds to that of the dominant stellar
population, i.e.\ the population that generates principally red HB stars.} 
there is no radial variation in the HB morphology
index.  However, outside the core radius, there are indeed relatively more
blue HB stars.  This is illustrated in Fig.\ 6 which shows cmds for the 
regions with r $\le$ 900 pix (i.e.\ $\sim$90$\arcsec$, upper panel) and 
r $\ge$ 900 pix (lower panel).
The average distance from the center of And I in these cmds is
$\sim$60$\arcsec$ for the
stars in the inner sample and $\sim$100$\arcsec$ for the outer sample.
It is apparent from these cmds that the outer region has a higher proportion
of blue HB stars.  Quantitatively, we have {\it i} = 0.109 $\pm$ 0.013 for the
inner region ({\it b} = 64, {\it r} = 523) and {\it i} = 0.195 $\pm$ 0.031 
for the outer region ({\it b} = 31, {\it r} = 128).  In both cases the error is
calculated assuming Poissonian statistics.
 
\begin{figure}
\figurenum{6}
\epsscale{0.80}
\plotone{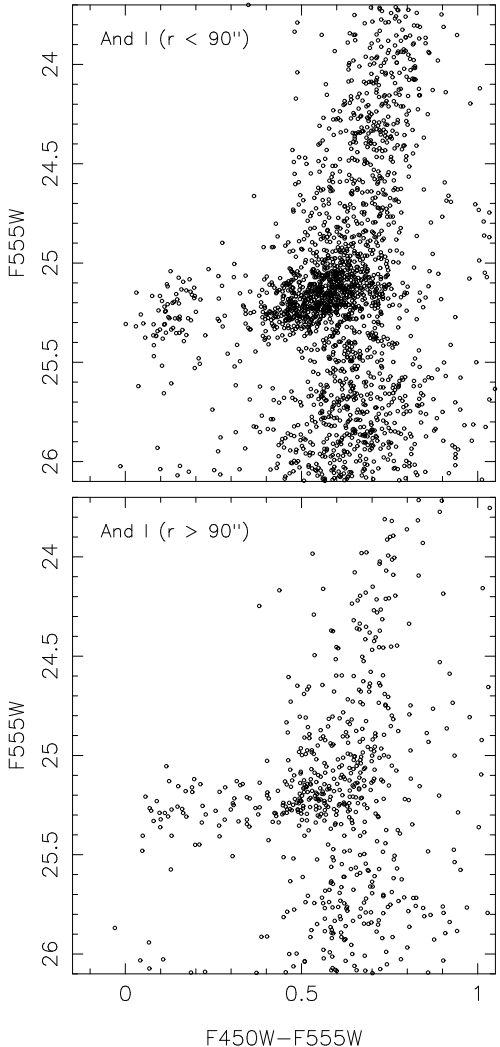}
\caption{And I color-magnitude diagrams 
emphasizing the change in
horizontal branch morphology with distance from the center of this dSph.  The
upper panel is for stars within 90$\arcsec$ of the center,
while the lower panel is for stars at radial distances of more than 
90$\arcsec$\@.  The relatively larger fraction of blue HB stars in the lower 
panel is evident.}
\end{figure}

How significant is this difference?  In comparing proportions, as we are doing
here, the relevant statistic (e.g. \markcite{AR90}Arnold 1990) is:
$$
T_2 = \frac{\hat p_1 - \hat p_2}{
  [\hat p(1-\hat p)(n_1^{-1} + n_2^{-1})]^{1/2}
}
$$
where, in our case, $\hat p_1$ and $\hat p_2$ are the values of {\it i} for
the inner and outer regions, respectively, $n_1$ and $n_2$ are the total
numbers of blue and red HB stars in each region, and $\hat p$ is the value
of {\it i} for the entire (inner + outer) sample.  Inserting the numbers into
this equation then yields $T_2$ = 2.88 which, since the statistic has a normal
distribution with mean zero and unit variance, implies that there is a less
than 1\% chance that the two proportions are drawn from the same underlying
population.
Alternatively, since the outer group is approximately 30\% of the
inner group in size, we can run simulations in which 3 from every 10 stars in
the inner group are selected and the value of the HB
morphology index calculated.  We find that for 120 independent samples 
of the same size as the outer data set,
the {\it largest} value of {\it i} was 0.175, 
still notably less than the
observed value of {\it i} for the outer group.  Thus, once again, it
appears that there is a less than 1\% probability that the observed
HB morphology difference arises by chance.

We conclude therefore that from inside to outside approximately one core
radius, the relative number
of blue HB stars increases by somewhat less than a factor of two.  In 
other words, it appears that the blue HB population of And I is more widely
dispersed than the red HB population.  
There does not, however, appear to be 
any other difference between the two groups of HB stars.  For example, the 
mean F555W magnitudes of the red HB stars are 25.21 and 25.22 for the inner
and outer regions, respectively, while for the blue HB stars, the 
corresponding values are 25.25 and 25.27.  The mean F450W--F555W colors are
also similar: 0.13 and 0.52 for the inner blue and red HB stars, and 
0.14 and 0.51 for the HB stars in the outer sample.

What change in fundamental characteristics of the stellar population of
And I could contribute to this observed gradient in HB morphology?  Ignoring
other possible (but not easily quantified) causes such as variable
mass-loss rates, etc, inspection of the HB calculations of 
Lee {\it et al.\ }(1994)
shows that first, at a mean abundance
near [Fe/H] = --1.5, a decrease in age of 3 to 4 Gyr is required to change
a blue HB into a red one, and second, at an age of $\sim$12 to 15 Gyr, 
an increase
in abundance of perhaps 0.6 dex is required for the same effect.  Suppose then
that the HB of And I is made up of two components which differ in age or
abundance by these amounts.  In the inner region the (younger or more 
metal-rich) red HB population comprises $\sim$86\% of the total while in the
outer region it is somewhat less dominant, contributing $\sim$75\% of the
total.  Could we detect these age or abundance changes via, for example,
the effect on the mean giant branch color?  If age is the distinguishing
characteristic then the answer is a definite ``no''; the lack of sensitivity
of giant branch color to age combined with the small relative number of the
``old'' stars means that the predicted difference in giant branch color is
negligible.  If however, an abundance difference of $\sim$0.6 dex is the
distinguishing characteristic, then the approximate doubling in size of the
contribution of the metal-poor component from the inner to the outer region
would generate a change in the mean B--V color of the upper giant branch
of $\sim$0.03 mag, with the color for the outer region being bluer.  In fact
no such color difference is seen in the actual data, though the small numbers
of stars on the upper giant branch in the outer region limit the precision of
this test.  We cannot therefore easily identify the underlying cause of
the HB morphology gradient; neither an abundance gradient nor a change in
the mean age (or a combination of both) can be positively ruled out.
Nevertheless, as we will see in Sect.\ 4, the existence of a 10 -- 15\%
population that is 0.5 -- 0.6 dex more metal-poor than the mean is unlikely
given the relatively narrow intrinsic abundance distribution in this galaxy.

\subsubsection{The Giant Branch Intrinsic Color Width}

In their study of the upper $\sim$two magnitudes of the And I giant branch,
MK90 pointed out that the observed color width in their cmd was considerably
larger than
that expected from their photometric errors.  Thus an intrinsic color range
was clearly present.  By analogy with the galactic dSph systems, where 
intrinsic
abundance spreads are well established (e.g.\ \markcite{NS93}Suntzeff 1993),
MK90 then concluded that the intrinsic color range on the And I giant branch
was due to a spread in abundance; they estimated that more than the central 
two-thirds of the distribution was contained in the abundance 
interval \mbox{--2.0 $\lesssim$ [Fe/H] $\lesssim$ --1.0}. 

We can also use our data to characterize this property of And I by 
investigating the giant branch color width at an appropriate luminosity.  
Ideally this luminosity would be below that of the horizontal branch where
contamination by AGB stars is not a concern.  However, at these fainter 
magnitudes, the relatively large size of the photometric errors as compared
to the observed color width precludes reliable analysis.  At brighter
magnitudes, for example 24.5 $\gtrsim$ F555W $\gtrsim$ 23.5, it is evident
from Fig.\ 3 that stars evolving away from the red horizontal branch onto the
AGB would compromise any analysis of the red giant branch color width.  
However,
based on the c-m diagrams of galactic globular clusters, it does 
appear that the AGB eventually either terminates or merges into the red
giant branch.  Thus the stars at higher luminosities can be used to constrain
the size of any intrinsic color width present.  We have chosen to study the
color width of the giant branch over the magnitude interval 23.5 $\ge$ F555W
$\ge$ 22.7 since at brighter magnitudes the giant branch turns over, becoming
almost horizontal ({\it cf.\ }Fig.\ 3) making it difficult to
easily interpret any color width.  The faint magnitude limit is set to minimize
the influence of AGB stars.

As a first step we fitted a low order polynomial to the stars
in the magnitude interval 24.2 $\ge$ F555W $\ge$ 22.4, excluding obvious
outliers.  Then for each star with 23.5 $\ge$ F555W $\ge$ 22.7, we determined 
the residual in F450W--F555W color from this mean giant branch at the F555W
magnitude of the star.  The distribution of these residuals for the 105 stars
considered And I giants is shown as a histogram in Fig.\ 7, together with
the residuals for the outlier stars that were excluded from the 
determination of the
mean giant branch.  The distribution for the assumed And I stars
appears to 
be reasonably uniform over a $\sim$0.2 mag full range, but statistically the
distribution
is not significantly different from a gaussian.  It can be
characterized
in a number of ways.  For example, the standard deviation $\sigma$ is 0.06 mag,
the inter-quartile range is 0.09 mag and the central two-thirds of the 
distribution
is contained in a residual range of 0.14 mag.  These values do not change 
substantially if the magnitude interval considered is altered by reasonable
($\sim$0.2 - 0.3 mag) amounts.  Similarly, if we divide the sample into two
groups consisting of the stars inside and outside $\sim$1 core radius, there
is no evidence for any difference in the respective residual distributions.

\begin{figure}[htb]
\figurenum{7}
\epsscale{1.00}
\plotone{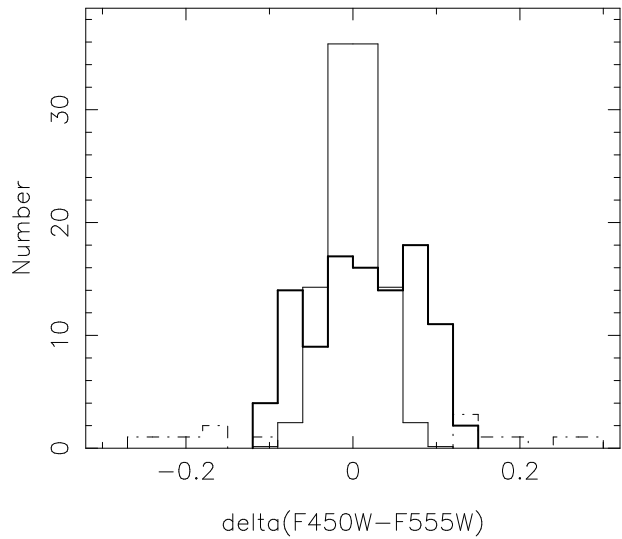}
\caption{Histogram of F450W--F555W residuals 
from the mean And~I giant branch for the magnitude interval
22.7 $\leq$ F555W $\leq$ 23.5.  The thick solid line is for the stars 
considered
likely members of And I while the dot-dash line is for those stars not 
considered And~I members.  The thin solid line is the distribution expected
on the basis of the instrumental errors alone, if And~I does not contain an
internal abundance range.}
\end{figure}

As discussed above, the photometric errors for these data have been determined
by comparing magnitudes and colors derived from the two independent sets of
observations.  For the magnitude interval considered here, we have 
$\sigma$(F450W--F555W) $\approx$ 0.02 mag, which is considerably smaller than
the observed standard deviation.  A gaussian distribution with this $\sigma$
and the same sample size is also plotted in Fig.\ 7 to emphasize the
difference between the observed distribution and the distribution expected 
on the basis of the photometric errors alone.  We emphasize that although
errors in the calibration and in the adopted aperture corrections,
for example, will affect
the zeropoint of the observed colors, the parameters characterizing the
distribution of residuals from the mean giant branch are unaltered by these
systematic effects.  Hence we conclude that the spread in color on the 
And I giant branch implied by Fig. 7 is dominated by an intrinsic range in
the colors of the And I stars ({\it cf.\ }MK90).  We defer, however, to 
section 3.4 a discussion of the corresponding abundance range in And I.
 
\subsubsection{A Population of Blue Stragglers?}

In the lower left portion of Fig.\ 3 a small population of faint (F555W 
$\gtrsim$ 26.0), blue (F450W--F555W $\lesssim$ 0.25) stars is visible.  At
F555W magnitudes between 26.0 and 26.4, the mean repeatibility error in the
F450W--F555W colors is less than 0.15 mag.  Thus it seems unlikely that all
of these faint blue stars could result from 3 or 4$\sigma$ errors in the 
photometry.  We note though that incompleteness in the F450W data, which
begins to be significant at perhaps F450W $\approx$ 26.8, means that we 
should not expect a symmetrical scatter in the F450W--F555W colors about the 
mean giant branch ridge line at these magnitudes.  For example, the relative
lack of stars in Fig.\ 3 with F555W $\gtrsim$ 26.0 and F450W--F555W
$\gtrsim$ 1.0 is probably a consequence of this effect.  Nevertheless, the 
fact that the faint blue stars in Fig.\ 3, like all the stars plotted, have 
been detected and measured {\it on both frame pairs} suggests strongly 
that these stars do represent a {\it bona fide} And I population.

What then are these faint blue stars?  There are two possibilities.  First,
the stars could be on or near the main sequence turnoff for an And I
population with
an age of perhaps 1.5 to 2.5 Gyr.  However, although we cannot completely
dismiss this possibility, we consider it unlikely since there is no other
evidence to suggest the presence of an intermediate-age (age $\lesssim$
10 Gyr) population in this dSph.  In particular, MK90 give 2$\sigma$ 
upper limits 
of 10\%, for the fraction of any And I population with ages between 0.5 and 
3 Gyr and 20\%, for a 3 to 10 Gyr population.  Similarly, Armandroff 
{\it et al.\ }(1993), using the data of MK90, estimate a 3 to 10 Gyr 
population fraction in And I of 10 $\pm$ 10\% while Armandroff (1994) 
reports that no luminous upper-AGB carbon stars have been detected in 
And I\@.  

The alternative explanation for these faint blue stars is that they are 
{\it blue stragglers} similar to the brightest examples seen in galactic
globular clusters and in some of the galactic dSph
galaxies (e.g. Ursa Minor, \markcite{OA85}Olszewski \& Aaronson 1985). 
Inspection of Fig.\ 10 of \markcite{FP92}Fusi Pecci
{\it et al.\ }(1992) for example, supports this interpretation.  In that 
figure, which shows a c-m diagram for a composite sample of 425 blue
stragglers in 21 galactic globular clusters, the brightest such stars have
(B--V)$_{0}$ colors between --0.10 and 0.35 and absolute visual magnitudes
between $\sim$1.5 and 2.0\@.  Anticipating the results of the next section,
these limits correspond to -0.05 $\lesssim$ B--V $\lesssim$ 0.40 and 
26.2 $\lesssim$ V $\lesssim$ 26.7 at And I\@.  These values are consistent
with the location of the And I faint blue stars in Fig. 4, though the brightest
of the And I candidate blue stragglers, at V $\approx$ 26.0 (M$_{V}$ $\approx$
1.3), are about 0.2 mag brighter than the brightest of the galactic 
globular cluster blue stragglers.  We do not, however, regard this as 
a serious concern.  If,
as discussed below, the predominantly red HB in And I indicates an age for the
majority of the population that is younger than the galactic globular clusters,
then the turnoff mass in And I will be correspondingly higher.  This in turn
will permit the formation of somewhat more massive, and therefore more 
luminous, blue straggler stars.

\subsection{The Distance of And I}

The reddening maps of \markcite{BH82}Burstein \& Heiles (1982) 
indicate 0.03 $\leq$ 
E(B--V) $\leq$ 0.06 for And I; we adopt E(B--V) = 0.04 $\pm$ 0.02 and 
thus $A_{V}$
= 3.2E(B--V) = 0.13 $\pm$ 0.06 mag.  In Fig.\ 4, the 89 stars with 25.0 $\leq$
V $\leq$ 25.4 and 0.0 $\leq$ B--V $\leq$ 0.35 yield a mean magnitude for the
horizontal branch of 25.25 $\pm$ 0.04\@.  The
statistical uncertainty in this value is small ($\sim$0.01 mag) but we have 
adopted $\pm$0.02 mag as the error in the mean aperture corrections and 
$\pm$0.03 mag as the uncertainty in the adopted F555W (V) magnitude zeropoint.
Walker (1995, {\it priv.\ comm.}) has indicated that the V magnitude 
zeropoint appears stable at this level over a four month interval that
encompasses the And I observations (see also, \markcite{R95}Ritchie 1995).  
To convert this $V_{HB}$ value
into a distance modulus for And I however, requires the adoption of a
horizontal branch luminosity calibration.  Such calibrations usually take the
form $M_{V}(RR) = a[Fe/H] + b$ though the values of the coefficients 
{\it a} and {\it b} remain contentious.  We follow
\markcite{DA90}Da Costa \& Armandroff (1990)
in adopting {\it a} = 0.17 and {\it b} = 0.82, which is the relation derived
from the horizontal branch models of \markcite{LD90}Lee, Demarque \& Zinn 
(1990) for a
helium abundance Y = 0.23\@.  This calibration is also the basis of the
distance scale that uses the I magnitude of the red giant branch tip 
in old stellar
populations (Da Costa \& Armandroff 1990; 
\markcite{LF93}Lee, Freedman \& Madore 1993a). 

Adopting a mean abundance of [Fe/H] = --1.45 $\pm$ 0.2 (see below),
then yields $(m-M)_{V}$ = 24.68 $\pm$ 0.05 and $(m-M)_{0}$ = 24.55 $\pm$ 0.08
for And I, corresponding to a distance of 810 $\pm$ 30 kpc.  This distance
agrees well with that given by MK90 (790 $\pm$ 60 kpc), which is based on the
I magnitude of the And I giant branch tip but which uses a calibration that 
assumes
$M_{V}$(RR) = +0.6 for all [Fe/H] $\leq$ --1.0.  Since, at the abundance of
And I derived here, $M_{V}$(RR) on our adopted scale is very close to 0.6, 
no adjustment of the MK90 result is required.  

To compare our distance to And I with that for M31 however,
requires an M31 modulus that is on the same distance scale.  There
are two appropriate determinations.  First, 
\markcite{PB88}Pritchet \& van den Bergh (1988)
have determined the mean magnitudes of RR Lyrae stars in an M31 halo field.
Adopting a mean abundance of [Fe/H] $\approx$ --1.0 for these stars 
(Pritchet \& van den Bergh 1988),
then yields an M31 distance modulus $(m-M)_{0}$ = 24.35 $\pm$ 0.15 on our
adopted scale\footnote{\markcite{vB95}van den Bergh (1995) gives a mean
V$_{0}$ of 25.04 for this same RR Lyrae sample, yielding $(m-M)_{0}$ = 
24.39 on our adopted distance scale.}.  Second, 
\markcite{MK86}Mould \& Kristian (1986) give I = 20.55 $\pm$ 0.15 mag
for the apparent I magnitude of the red giant branch tip in their M31 halo
field.  Assuming a reddening of E(B--V) = 0.08 $\pm$ 0.02 for this field and 
noting that $M_{I}(RGB_{tip}) \approx$ --4.05 $\pm$ 0.10 for old metal-poor 
red 
giants (Da Costa \& Armandroff 1990), then yields $(m-M)_{0}$ = 24.45 $\pm$ 
0.20 for M31 on our adopted distance scale.  Together these
determinations suggest $(m-M)_{0}$ = 24.40 $\pm$ 0.13 for M31 or a distance of
760 $\pm$ 45 kpc.  Thus, on the basis of these data, And I is
apparently some 50 $\pm$ 50 kpc
beyond M31 along the line-of-sight.  

The recent HST results of \markcite{AG96}Ajhar {\it et al.\ }(1996) however,
give a somewhat contradictory picture.  Ajhar {\it et al.\ }(1996) give
$V_{HB}$ values for three globular clusters and two field regions in M31 based
on WFPC2 images in the F555W and F814W filters.
Considering first the two metal-poor globular clusters (K105 and K219) where
no adjustment is necessary to convert $V_{HB}$ to $V_{RR}$, and noting that
we adopt a reddening E(B--V) = 0.045 $\pm$ 0.02 for K219, the mean $V_{HB}$ 
values of Ajhar {\it et al.\ }(1996) lead to distance moduli of 
24.75 $\pm$ 0.08 and
24.65 $\pm$ 0.07, respectively, on our adopted distance scale.  The 
uncertainties are those given by Ajhar {\it et al.\ }(1996) for their
$V_{0}(HB)$ values.  For the metal-rich cluster K58, we choose to adjust
$V_{HB}$ by the same amount as used by us for 47 Tuc (0.15 mag) leading to a
modulus of $(m-M)_{0}$ = 24.55 $\pm$ 0.11 for this cluster.  Together these
three clusters then imply an M31 modulus on our adopted scale
of 24.65 $\pm$ 0.1 corresponding to a
distance of 850 $\pm$ 40 kpc.  In this situation, And I is then apparently
40 $\pm$ 50 kpc in front of M31.  The two field regions considered in
Ajhar {\it et al.\ }(1996), which are near the clusters K58 and K108, are
undoubtedly contaminated by a younger population from M31's outer disk.  It
is therefore not straightforward to convert the observed mean magnitude for
the red horizontal branch (red clump?) to an equivalent $V_{RR}$.  
Nevertheless,
the Ajhar {\it et al.\ }(1996) data for these fields, taking a $V_{HB}$ to
$V_{RR}$ correction of 0.2 mag and assuming a mean abundance of [Fe/H] 
$\approx$ --0.3 (Ajhar {\it et al.\ }1996) also imply an M31 distance modulus
of approximately 24.60 on our adopted scale.  

Given these results then, it is apparent that we cannot at the moment precisely
determine the relative line-of-sight distance between M31 and And I; a value 
of 0 $\pm$ 70 kpc would seem to be the best estimate.  Clearly, an accurate
determination of $V_{HB}$ from a sizeable sample of M31 halo field blue
horizontal branch and RR Lyrae stars would be the best way to reduce
the uncertainty in this relative line-of-sight distance 
determination\footnote{In a recent preprint, 
\markcite{FP96}Fusi Pecci {\it et al.\ }(1996) have reanalyzed the 
Ajhar {\it et al.\ }(1996) data, combining the results with similar data 
for an additional four M31 globular clusters.  On our adopted scale, and with
our adopted correction to $V_{HB}$ for metal-rich red HB clusters (0.15 mag,
{\it cf.\ }0.08 mag in Fusi Pecci {\it et al.\ }1996),
the mean modulus for the eight clusters is $(m-M)_{0}$ = 24.64 $\pm$ 0.05, 
corresponding to a distance of 850 $\pm$ 20 kpc.  These values agree very well
with those given by Ajhar {\it et al.\ }(1996), and again indicate that And I 
lies 40 $\pm$ 40 kpc in front of M31.  Thus our adoption of 0 $\pm$ 70 kpc 
for the And I/M31 line of sight distance is unaltered by these new data.}.

At an M31 distance of between 760 and 850 kpc, the projected distance of 
And I from the center of M31 is $\sim$45 kpc, so that, given the above
results, And I lies at a true 
distance of between $\sim$45 and $\sim$85 kpc from 
the center of M31.  The lower limit is considerably smaller than the 
galactocentric distances of the nearer dSph companions to the Galaxy,
such as Ursa Minor ($R_{gc} \approx$ 70 kpc), Draco and Sculptor 
(both $R_{gc} \approx$ 80 kpc) though it does exceed the galactocentric 
distance of the
Sagittarius system, which has R$_{gc}$ of $\sim$16 kpc.  However, the Sgr
dSph is clearly
being strongly affected by the tidal field of the Galaxy while And I has a
smooth undistorted surface brightness profile (Caldwell {\it et al.\ }1992).  
This suggests a possible alternative approach to estimating the true 
distance of And I from the center of M31.

The surface brightness profile of Caldwell {\it et al.\ }(1992) yields
a ``tidal'' radius for And I of 2.8 $\pm$ 0.3 kpc from a King (1966) 
model fit.  Further, Caldwell {\it et al.\ }(1992) also give a total
integrated magnitude M$_{V}$ of --11.7 for And I\@. If we then assume
a mass-to-light ratio M/L$_{V}$ of $\sim$10 for And I 
({\it cf.\ }\markcite{MM93}Mateo {\it et al.\ }1993, Fig.\ 8
and \markcite{PC93}Peterson \& Caldwell 1993, Fig.\ 2), we can then
apply the relation of 
\markcite{KI62}King (1962, equation 12) to 
generate predicted tidal radii for various assumed And I perigalactic 
distances.  
If we take a M31 mass of $5 \times 10^{11} M_{\sun}$, then for a 
perigalactic distance of 45 kpc, the predicted tidal radius is 1.3 kpc, a
value more than a factor of two smaller than the observed limiting radius.
For a perigalactic distance of 90 kpc however, the predicted tidal radius is
$\sim$2.5 kpc in better accord with the Caldwell {\it et al.\ }(1992) 
observations.
Consequently, although there are obviously a number of uncertain factors in
this analysis, such as the assumed M/L for And I, the mass of M31 and more
significantly, whether
the observed limiting radius of And I is actually set by M31's tidal field 
({\it cf.\ }\markcite{PS85}Seitzer 1985), 
the results do suggest that the true distance of And I from the center
of M31 is nearer the upper limit suggested above (i.e. R $\sim$ 85 kpc) than
the lower.
 
\subsection{The Abundance of And I}

The determination of an abundance estimate for And I from these data requires
a comparison with the giant branches of galactic globular clusters of
known abundance.  Since there are no such giant branches available in the
WFPC2 (F555W, F450W--F555W) system, we are forced to rely on the 
transformations listed by H95b to convert our data to the standard B,V 
system.  The F555W filter is one of the primary photometric filters so that
it is well calibrated (H95b).  Similarly, the F555W to V transformation has
only a small color dependence, so that we do not expect this transformation
to introduce any significant uncertainties.  On the other hand, the F450W
filter is not one of the primary filters and its calibration is determined
synthetically (H95b).  This requires accurate modelling of the overall system
response; H95b suggest an uncertainty of $\sim$2\% in this approach.  Further,
the transformation of the F450W--F555W colors to standard B--V is also 
based on a
synthetic process, via the use of a library of stellar spectrophotometry 
(H95b).
This library (e.g.\ \markcite{GS83}Gunn \& Stryker 1983) contains 
principally bright solar neighborhood dwarfs and giants of presumably
approximately solar abundance.  Thus, given the large color coefficients in the
F450W--F555W to B--V transformation (F450W--F555W $\approx$ 0.71(B--V) + 
0.036(B--V)$^{2}$, H95b) its applicability to lower gravity metal-poor 
globular cluster giants is not clear-cut.  We have therefore begun a program
of observing standard globular clusters with ground-based telescopes using
F450W and F555W (and standard B, V) filters together with a CCD whose response
is similar to that of the WFPC2 devices.  However, this program 
is not yet complete
and so for the present we shall assume the applicability of
the H95b transformation for F450W--F555W keeping in mind the possibility 
of both zeropoint and systematic effects.

In Fig.\ 8 we again present the And I photometry, transformed to the 
(V, B--V) system
using the H95b transformations, together with the giant
branches for the galactic globular clusters M68 (\markcite{W94}Walker 1994), 
M55 (\markcite{L77}Lee 1977),
NGC 6752 (\markcite{CS73}Cannon \& Stobie 1973), 
NGC 362 (\markcite{H82}Harris 1982) and 47 Tuc 
(\markcite{H87}Hesser {\it et al.} 1987).  
The adopted reddenings, V(HB) values and 
abundances of these clusters are given in Table 2\@.  These giant branches
were initially placed on the And I data using the ($M_{V}(HB)$, [Fe/H]) 
relation
described above, an apparent modulus $(m-M)_{V}$ = 24.68 and a reddening
E(B--V) = 0.04 mag.  However, with this reddening, the agreement between the
observations and the standard globular cluster giant branches was not
satisfactory, particularly for the region of the And I cmd fainter
than the horizontal branch.  As is shown in the figure, a more satisfactory fit
is achieved if the globular cluster giant branches are reddened by a 
further 0.05 mag in B--V, with no change in the V magnitudes.  Given the
uncertainty in the zeropoint of the F450W calibration and the uncertainty in
the zeropoint of our photometry (e.g. $\sim\pm$0.02 mag is the error in the 
adopted mean aperture corrections), this requirement for an additional 
color shift does not seem unreasonable\footnote{We have recently acquired B, V
CCD frames of And I using the WIYN telescope.  When reduced, these data will
clarify the situation regarding the zeropoints of the photometry
shown in Figs.\ 4 and 8\@.}.

\begin{figure}[htb]
\figurenum{}
\epsscale{1.00}
\plotone{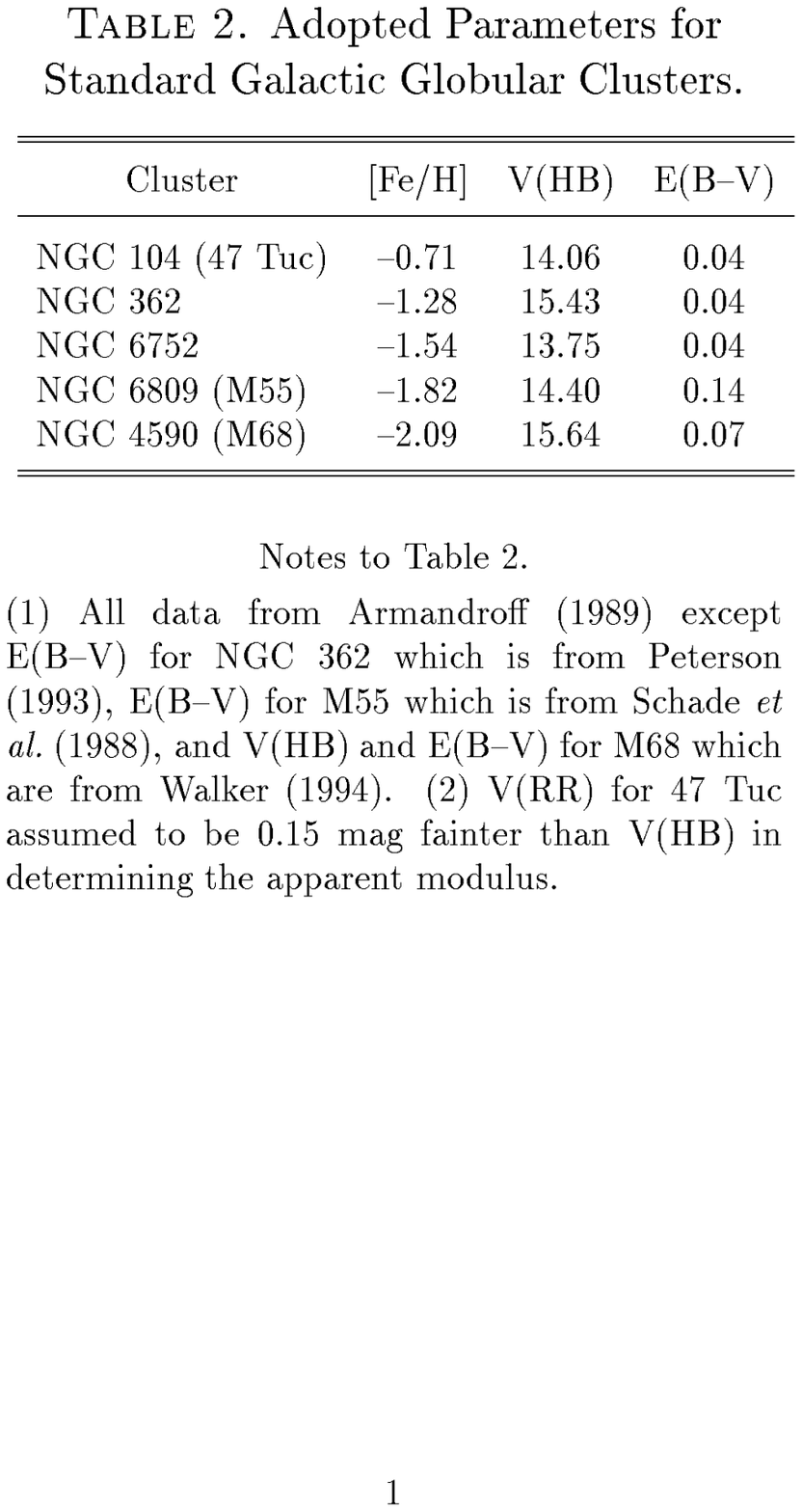}
\end{figure}

\begin{figure}[htb]
\figurenum{8}
\epsscale{1.00}
\plotone{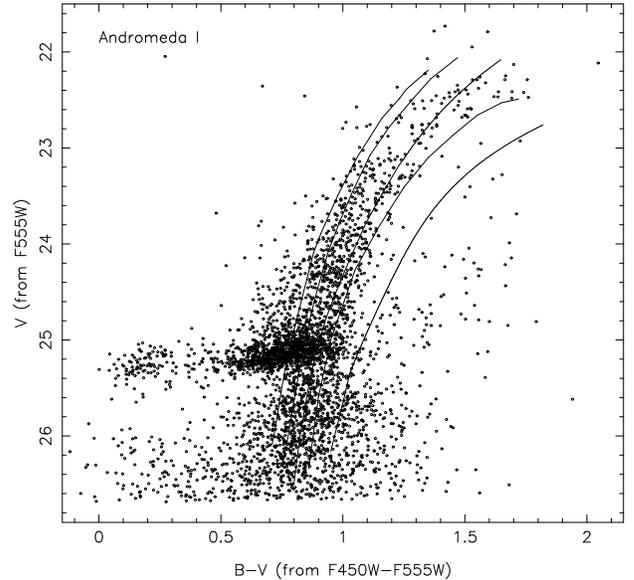}
\caption{The (V, B--V) color-magnitude diagram for And~I 
on which the giant branches for the standard galactic globular clusters 
M68 ([Fe/H] = --2.09), M55 (--1.82), NGC 6752 (--1.54), NGC 362 (--1.28)
and 47 Tuc (--0.71) have been superposed.  The adopted parameters for
the galactic globular clusters are given in Table 2\@.  The giant branches
have been placed on the And~I data assuming the ($M_{V}(HB)$, [Fe/H]) relation
of Lee {\it et al.\ }(1990) for Y = 0.23, an apparent distance modulus
$(m-M)_{V}$ = 24.68, and a
reddening E(B--V) = 0.04 mag.  The giant branches were then shifted a further
0.05 mag to the red (with no change in the V magnitudes) to improve the
overall fit.}
\end{figure}

We can now proceed to derive an estimate for the mean abundance of And I; an
estimate of
the size of the abundance spread implied by the intrinsic color width 
discussed in Sect.\ 3.1.3 above will be derived in the next section.  
The mean abundance estimation can be carried
out at two possible V magnitude intervals: either at magnitudes below 
the horizontal
branch where contamination from stars evolving from the HB is not a concern,
or at relatively bright magnitudes where, at least in galactic globular 
clusters, the AGB is no longer visible having either merged into the red giant
branch or terminated.  The fainter magnitude interval 
has the advantage of more stars
but the photometric errors are larger, as is the degree of contamination from
non-member objects (galactic foreground stars, M31 halo stars and 
background galaxies).  The sensitivity to abundance is also reduced.
Nevertheless, the mean B--V color for the 141 stars with 25.5 $\le$ V $\le$
25.7 and 0.66 $\le$ B--V $\le$ 1.02 in Fig.\ 8 is 0.86 $\pm$ 0.01, where the
uncertainty given is the standard error of the mean.  For the 166 stars
with 25.7 $\le$ V $\le$ 25.9 and the same color limits, the mean B--V color
is also 0.86 $\pm$ 0.01.  The B--V color limits were chosen by inspecting 
histograms of the F450W--F555W color distributions for the equivalent F555W
magnitude intervals in conjunction with the expected color error
distributions.  This allowed us to set limits which would comfortably
include most possible members while excluding the majority of contaminating
objects.  Linear least squares fits to the B--V colors of the five standard
globular cluster giant branches at V = 25.6 and V = 25.8 then enable the
mean B--V colors to be converted into abundance estimates.  The resulting
values are [Fe/H] = --1.50 $\pm$ 0.20 and [Fe/H] = --1.42 $\pm$ 0.24 for the
brighter and fainter V magnitude intervals, respectively.  The abundance
uncertainties given include the error from the calibration, the
statistical uncertainty in the observed mean color, and an adopted systematic
uncertainty of $\pm$0.03 mag in the mean color.  Together these values
indicate an abundance for And I of [Fe/H] = --1.45 $\pm$ 0.20 dex.  This value
is in good accord with that, [Fe/H] = --1.4 $\pm$ 0.2, determined by MK90
from the mean (V--I)$_{0}$ color of And I's upper giant branch.

Inspection of the color-magnitude diagrams for the standard globular
clusters indicates that AGB stars cease to be readily distinguishable from
red giant branch stars brighter than M$_{V} \approx$ --1.5, corresponding to
V $\approx$ 23.2 in And I\@.  We have therefore derived a second estimate for
the mean abundance of And I by considering the mean (B--V) color of
the stars in the magnitude intervals 22.5 $\le$ V $\le$ 22.7, 22.7 $\le$ V
$\le$ 22.9, 22.9 $\le$ V $\le$ 23.1, 23.1 $\le$ V $\le$ 23.3 and 23.3 $\le$
V $\le$ 23.5.  Excluding a few obvious outliers, there are 23, 17, 13, 40 and
42 stars in these bins and, again using the standard globular cluster giant
branches as calibration, the resulting mean abundances range from [Fe/H] =
--1.73 to --1.64 dex, with uncertainties of order 0.25 dex.  At these
magnitudes, the abundance uncertainty is dominated by the uncertainty in the 
calibration
rather than any systematic (assumed to be $\pm$0.03 mag) or statistical
uncertainty in the mean B--V colors, though these have been included in the
error calculation.  Further, there is no
systematic trend with luminosity in these upper giant branch
abundance estimates, supporting our
assertion that AGB stars have not biased the abundance estimates.  We thus 
determine a mean abundance of [Fe/H] = --1.67 $\pm$ 0.25 dex for And I from
the upper giant branch stars.

This abundance is some 0.25 dex or so more metal-poor than that derived from 
the lower giant branch.  It also differs by about the same amount from the
abundance determination of MK90, which is based on the mean V--I colors of 
stars in a similar magnitude interval on the upper giant branch.  Since 
we do not believe that AGB stars have 
significantly biased our result for the upper giant branch, the origin of
this difference presumably lies with the F450W--F555W to B--V transformation.
As noted above, the F450W--F555W to B--V transformation of H95b used here
is derived synthetically from a spectrophotometric library of approximately
solar abundance stars.  It is therefore not surprising, given the difference 
in bandpass and effective wavelength between the F450W and B filters, that the
differences in blanketing between solar abundance stars and metal-poor red
giants can give rise to systematic errors.  These effects are likely to be
most marked for cooler redder stars and for that reason we prefer the And I
abundance derived from the lower giant branch.  Interestingly however,
if we do not make the additional 0.05 mag shift in B--V color described above,
then the abundance derived from the upper giant branch stars, [Fe/H] = --1.55
$\pm$ 0.25 is in better accord with the value derived from the lower giant
branch with the color shift applied (without the color shift, the 
lower giant branch yields an abundance of [Fe/H] = --1.20 $\pm$ 0.25 dex).  We
therefore adopt [Fe/H] = -1.45 $\pm$ 0.2 dex as our best estimate of the
mean abundance of And I, pending further investigation of the F450W--F555W
to B--V transformation.
  
\subsection{The Abundance Spread in And I}

In section 3.1.3 we demonstrated that in the magnitude range 23.5 $\ge$ F555W
$\ge$ 22.7, the And I giant branch possesses an intrinsic color spread
characterized by, for example, an (error corrected) standard deviation 
in the color residual
from the mean giant branch of $\sim$0.06 mag.  We can now use the H95b
transformations to B--V and the abundance calibrations derived above
to convert these intrinsic F450W--F555W color residual measures to
residual measures in abundance.  We note in particular that since the 
abundance 
calibration is linear, the resulting abundance residuals do not depend on
the mean abundance of And I, i.e.\ the validity of the 0.05 mag additional
shift in B--V plays no role in the determination of the abundance residuals.
Applying the transformations and the abundance calibrations then yields the
following: $\sigma$([Fe/H]) = 0.20 dex, inter-quartile range 0.30 dex,
central two-thirds of the sample abundance range 0.45 dex, and full
abundance range for this sample of $\sim$0.6 dex.  

The abundance range for the central two-thirds of the abundance distribution
determined
here (0.45 dex) is considerably smaller than that ($\sim$1 dex) found by 
MK90.  We
attribute this difference principally to the fact that the superior resolution
of HST and WFPC2 has produced colors and magnitudes with significantly smaller
photometric errors over a similar magnitude interval to that used by
MK90, resulting in less uncertainty in the derived abundance
range.  Specifically, MK90 determined the abundance range for the
central two-thirds of their abundance distribution
from an {\it observed} color range of
$\sim$0.4 in V--I with errors
$\sigma_{V-I} \approx$ 0.13 mag, dominated by image crowding.  Our value comes
from an observed F450W--F555W color range of 0.14 mag but with errors
$\sigma_{F450W-F555W}$ = 0.02 mag; the effect of the errors is thus
considerably smaller in the present case. 

Internal abundance spreads are well established in the galactic dSphs through
both photometric and spectroscopic techniques.  For example, Suntzeff (1993)
lists values of $\sigma$([Fe/H]) for five systems ranging from $\sim$0.2
to $\sim$0.3 dex.  Similarly, Armandroff {\it et al.\ }(1993)
have photometrically determined 0.16 $\le$ $\sigma$([Fe/H]) $\le$ 0.24 for
the M31 dSph companion And III, while 
\markcite{KM93}K\"{o}nig {\it et al.\ }(1993)
claim a large abundance dispersion, $\sigma$([Fe/H]) $\approx$ 0.43, in the 
And II dSph galaxy.  It is then apparent that the present result
of $\sigma$([Fe/H]) = 0.20 dex for And I is one of the lower internal 
abundance 
dispersions measured in a dSph galaxy. 

Naively one might expect that the abundance spread in a dSph galaxy would be
larger in the higher mean abundance, higher luminosity (these two quantities
are well correlated, e.g.\ Caldwell {\it et al.\ }1992) systems, as more
generations of stellar processing produce not only a higher mean abundance
but also a higher abundance dispersion (see, for example, eq.\ 5 of 
\markcite{JM84}Mould 1984).  But this does not seem to be the 
case: among the And systems we see that And I and And III have similar values
of $\sigma$([Fe/H]) while that of And II is apparently considerably larger,
yet according to Caldwell {\it et al.\ }(1992), And I and And II have identical
luminosities which exceed that of And III by approximately 1.5 magnitudes. 
Similarly, \markcite{SM93}Suntzeff {\it et al.\ }(1993) have spectroscopically
determined $\sigma$([Fe/H]) = 0.19 $\pm$ 0.02 for the Sextans dSph, a value
closely similar to that derived here for And I, yet Sextans is a dSph system
that is $\sim$1.7 magnitudes or more fainter than And I\@.  Thus on 
the basis of presently available data, it is hard to make a compelling case 
that $\sigma$([Fe/H]) increases at all with luminosity among the M31 and 
galactic dSph companions.  Clearly there is considerable scope for progress
in this subject area, with large samples of spectroscopic abundance
determinations, where practical, being the key ingredient.  It may then be 
possible to use a ($\sigma$([Fe/H]), [Fe/H] or M$_{V}$) correlation, or lack
thereof, to constrain the formation and chemical evolution of these 
dwarf galaxies.

\section{Discussion}

\subsection{And I and the Outer Halo of M31}

The principal result of this study has been to reveal that the dSph galaxy
And I, which lies in the outer halo of M31, has a predominantly red
horizontal branch morphology.  When combined with the metal abundance estimate
[Fe/H] = --1.45 $\pm$ 0.2 dex, this red HB morphology indicates that And I
can be classified as a ``second parameter'' system; i.e.\ it possesses a
redder HB than the majority of galactic globular clusters of comparable 
abundance.  And I is thus similar to many of the galactic dSph galaxies
which are also classified as second parameter objects.

In the Galaxy, second parameter systems (both globular clusters and dSphs) are
most common in the outer regions of the halo, while they are conspicuously
absent from the inner halo region.  {\it If} the second parameter effect is
primarily driven by age differences, then this variation in the frequency of
occurrence of the second parameter effect with galactocentric distance 
indicates that the 
mean age is smaller, and the age range larger, in the outer galactic halo 
(e.g.\ Lee {\it et al.\ }1994; but
see \markcite{HR96}Richer {\it et al.\ }1996 for an alternative view).  This
age variation interpretation is then one of the
cornerstones of the halo formation model advocated initially by Searle \&
Zinn (1978; see also \markcite{ZR93}Zinn 1993b).  In this model the outer
halo is built up by mergers and accretions from a number of smaller 
independently evolving satellite galaxies.

While it is obviously premature to derive any strong conclusions from the
observations of a single object, the existence of a second parameter system
in the outer halo of M31, namely And I, is consistent with the hypothesis
that the M31 outer halo formed by the same drawn-out chaotic process as
postulated for the outer halo of the Galaxy.  Certainly the lack of any
substantial abundance gradient in the M31 globular cluster system
\markcite{JH93}(Huchra 1993) supports this contention\footnote{In the Galaxy,
the lack of any radial abundance gradient in the halo globular cluster
system is a second cornerstone of the postulated merger-accretion halo 
formation process.}
though the most distant
clusters studied are only at 25 - 30 kpc in projection from the center of
M31 ({\it cf.\ }$\sim$45 kpc for And I and up to $\sim$100 kpc in the
Galaxy).  Similarly, the HB morphologies 
of the three metal-poor ([Fe/H] $\sim$ --1.8 to --1.5 dex) M31
globular clusters studied by Fusi Pecci {\it et al.\ }(1996), which lie at
projected distances of between $\sim$15 and 20 kpc from the center of M31,
do not show any obvious indication of second parameter characteristics.  
When combined with our result for And I, this is consistent with
the diversity of HB morphologies expected as a consequence
of the merger-accretion halo formation process.  On the other hand, the 
mean abundance of the M31 halo population ($<$[Fe/H]$>$ $\approx$ --0.6,
Durrell {\it et al.\ }1994) is considerably higher than that for the Galaxy.
The M31 abundance would appear to be difficult to generate from the 
disruption of satellite galaxies unless the systems were predominantly more 
massive in the case of M31
than was the case for the Galaxy.  It is worth noting though
that dSphs with metal abundances comparable to the M31 halo mean abundance 
do exist.  For example, the 
Sagittarius dSph galaxy, which is now merging into the halo of the Galaxy
(\markcite{IG94}Ibata {\it et al.\ }1994), has a high mean abundance; estimates
range from $<$[Fe/H]$>$ = --1.0 $\pm$ 0.3 (Ibata {\it et al.\ }1994) to
$<$[Fe/H]$>$ = --0.5 $\pm$ 0.1 (\markcite{SA95}Sarajedini \& Layden 1995).
Clearly then the comparatively high mean abundance for the halo field
population in M31 is
not a fatal objection to the merger-accretion halo formation process.

We are then in reality just beginning to explore the degree of similarity
between the halo formation processes for the Galaxy and M31.  Our present
results for And I provide some support for the hypothesis that these halo
formation processes were indeed similar, but many more
objects in the outer halo of M31 with well determined HB morphologies and 
abundances are required before this issue can be settled.

\subsection{The Age(s?) of And I}

Turning now to And I itself, we first seek an estimate for the
age of the majority population in this dSph.  Over the past decade or more,
studies reaching the main sequence turnoff in galactic dSphs have indicated
that the age of the majority population varies substantially from dSph to
dSph.  Ursa Minor, for example, is apparently as old as the oldest galactic
globular clusters (Olszewski \& Aaronson 1985) while systems such as Carina
(\markcite{MA83}Mould \& Aaronson 1983, \markcite{KM90}Mighell 1990,
\markcite{TS94}Smecker-Hane {\it et al.\ }1994) 
and Leo I (\markcite{ML93}Lee {\it et al.\ }1993b,
\markcite{MO94}Mateo {\it et al.\ }1994) are dominated by stars of
intermediate ($\sim$2 to 8 Gyr) age.  For And I, observations reaching the
main sequence turnoff (V $\approx$ 28.7 for a population as old as most 
galactic globular clusters) are not currently practical and
so we must seek alternate methods to provide an age estimate.

The most appropriate method is that of Sarajedini {\it et al.\ }(1995,
hereafter SLL95; see also \markcite{DH91}Hatzidimitriou 1991) which is based
on the difference between the mean color of the giant branch at the level
of the horizontal branch and that of the red HB.  At fixed abundance, this
color difference increases with increasing age.  For And I, the lack of a clear
separation between the red end of the HB and the giant branch
({\it cf.\ }Figs.\ 3 \& 4) complicates the application of the SLL95 method.  
We have
therefore proceeded as follows.  First, to determine the mean color of the 
giant branch at the level of the horizontal branch (V = 25.25), we initially
determined the mean giant branch color for the magnitude interval
25.4 $\leq$ V $\leq$ 25.8, which lies just below the horizontal
branch.  We then employed the giant branches of the standard globular
clusters NGC 362 and NGC 6752, which bracket And I's mean abundance, to
correct this mean color to a value appropriate for the magnitude of the
horizontal branch.  This process then yields (B--V)$_{g}$ = 0.88 $\pm$ 0.02 
mag.  
Next, to determine the mean color of the red horizontal branch, we scaled the
B--V color histogram for the giant branch in the magnitude interval
25.4 $\leq$ V $\leq$ 25.8 by the relative number of stars on the red side
of the equivalent histogram for the interval 25.0 $\leq$ V $\leq$ 25.4,
shifted it to the red by 0.03 mag to compensate for the giant branch slope,
and then subtracted it from the distribution at the brighter magnitude 
interval.
This generates a color distribution which is predominantly red HB stars.  The
mean color of this distribution is $<$(B--V)$_{HB}$$>$ = 0.70 $\pm$ 0.02,
where the error given incorporates both the uncertainty in the adopted scaling
and in the color interval used to calculate the mean.  These two mean values
then give $d_{B-V}$ = 0.18 $\pm$ 0.03 mag.  

Combining this determination with our adopted And I metal abundance of 
[Fe/H] = --1.45 $\pm$ 0.2, Fig.\ 4 of SLL95 then yields an age for And I's 
majority population of 9.5 $\pm$ 2.5 Gyr.  This age is younger than the ages
of most of the galactic globular clusters, indeed it is even some 3 Gyr or so 
younger than the ages derived via this method for classic second parameter
clusters such as NGC 362 (SLL95).  It is not, however, inconsistent 
with the large spread of mean ages seen among the galactic dSphs.  
Nevertheless, it is reasonable to investigate the reliability of this age
determination.  We note first that
the lack of any observed upper-AGB population in And I
suggests strongly that there is no large population present
with ages less than $\sim$10 Gyr, which is consistent with the 
above result.  Second, the recent HST based results of Mighell \& Rich (1996)
for Leo II provide an additional consistency check.

These authors present observations that reach fainter than the
main sequence turnoff in this dSph, from which they conclude that the main
stellar population in Leo II is 8 -- 10 Gyr old.  As noted above, this 
galactic dSph has a similar HB morphology to And I though its metal abundance
is somewhat lower (Mighell \& Rich (1996) give [Fe/H] = --1.6 $\pm$ 0.25).
Given this main sequence turnoff age determination, we can then apply the
SLL95 method to the Leo II red HB and giant branch to see if consistent
results are achieved.  Mighell \& Rich (1996) observed in the (F555W, F814W)
system and they give V--I = 0.828 for the median color of the red HB and
V--I = 0.970 for the color of the giant branch at the level of the
horizontal branch (note that these are observed, rather than reddening 
corrected values).  The method of SLL95, however, requires B--V colors and
so we have converted these V--I values to B--V using a linear relation
between V--I and B--V that is based on the V--I photometry of Da Costa
\& Armandroff (1990) and the B--V photometry of Cannon \& Stobie (1973) for
giants in the globular cluster NGC 6752.  This cluster has 
both similar reddening
and similar abundance to Leo II\@.  The transformation gives 
(B--V)$_{g}$ = 0.786 and $<$(B--V)$_{HB}$$>$ = 0.611, yielding
$d_{B-V}$ = 0.175\footnote{This value is slightly larger than that given by
Mighell \& Rich (1996), who used a quadratic (B--V, V--I) transformation
derived from colors given in the Revised Yale Isochrones.}.
Adopting [Fe/H] = --1.6 for the abundance of Leo II, this value of $d_{B-V}$
yields an age of 8.5 Gyr from Fig.\ 4 of SLL95.  This value is in excellent
agreement with the results of Mighell \& Rich (1996) who give, based on a
number of methods principally based on the turnoff luminosity, a mean age of
9.1 $\pm$ 0.8 Gyr for Leo II's dominant stellar population.

We conclude therefore that the SLL95 method as applied to our And I
observations is relatively reliable and that therefore the bulk of the
stellar population in this M31 dSph companion is approximately 10 Gyr 
old.  In comparison with the galactic dSphs, this result 
indicates that And I is comparable to Leo II in age though it is definitely
older than systems such as Carina and Leo I\@.  It is probably somewhat
younger than Sculptor, however, for which \markcite{GD84}Da Costa (1984) 
has found
an age some 2 to 3 Gyr younger than the majority of galactic globular
clusters (in all these cases the age refers to the bulk of the stellar
population). 

One question remains.  In Sections 3.1.3 and 3.4 we pointed out that And I
has an intrinsic abundance dispersion.  We have also shown that And I
possesses a small (10 -- 15\%) population represented by blue HB and RR Lyrae
variable stars.  Is it possible that these blue HB stars and RR Lyrae variables
result from the metal-poor tail of the abundance distribution at approximately
constant age, or is it necessary to invoke a significant age difference
between these stars and the bulk of the And I population?  Again employing
the HB models of Lee {\it et al.\ }(1994), it appears that at constant age,
an abundance decrease of approximately 0.5 dex from [Fe/H] $\approx$ --1.5
is needed to generate a reasonably blue HB (we assume (B--R)/(B+V+R) = 0.5).
We therefore need to establish if the abundance distribution in And I is
compatible with a $\sim$10\% population that is approximately 0.5 dex more
metal-poor than the mean.  The results of Sect.\ 3.4 suggest that this is
unlikely.  The required abundance difference exceeds 2$\sigma$ and for a
gaussian distribution, only $\sim$2\% of the population is beyond that limit.
Similarly, the size of the abundance range that contains 2/3rds of the sample,
and the total observed abundance range, both suggest that a 10\% sample
containing the most metal-poor stars will differ from the mean abundance
by only $\sim$0.25 dex.  This is not sufficient to generate a sufficiently
blue horizontal branch.  We conclude therefore that the blue HB and RR Lyrae
stars are more readily explained as coming from a population that is somewhat
older, by perhaps 3 Gyr or more, than the bulk of the And I population.  This
population may therefore be of comparable age to the globular clusters
of the inner galactic halo.

There is, however, a possible caveat that should not go 
unmentioned.  In Sect.\ 3.3 we noted that the mean abundance derived from the
upper giant branch was approximately 0.2 to 0.25 dex more metal-poor than
our preferred value.  If this more metal-poor mean abundance was to be
substantiated, then the low metal abundance tail of the
abundance distribution could generate blue HB and RR Lyrae variable stars
without requiring any substantial age difference.  However, we do not believe
this is the case.  In addition to the arguments cited in Sect.\ 3.3
supporting the adoption of the higher abundance, we note that if the lower
mean abundance for And I was correct, then the age derived from the 
$d_{B-V}$ value would be $\sim$8 Gyr.  Such an age for the bulk of the
stellar population in And I is inconsistent with the results that limit the
intermediate-age (3 -- 10 Gyr) population fraction in And I to 
$\sim$10 $\pm$ 10\% (e.g.\ Armandroff {\it et al.\ }1993, Armandroff 1994).

Consequently, we can restate our conclusion that the most reasonable
explanation for the existence of blue HB and RR Lyrae variable
stars in And I is that they represent a (small) population of stars that
formed perhaps 3 Gyr or more before the bulk of the stellar population in
And I\@.  Such a result implies that, just as is the case for many of the 
galactic dSphs, star formation in And I occurred over an extended interval.
The driving mechanism(s) behind the extended and diverse star 
formation histories of the Local Group dSph systems remains one of the 
most outstanding unsolved problems in this branch of Astronomy. 

\subsection{Horizontal Branch Morphology Radial Gradients}

A further property of And I that is worthy of additional comment is that
established in section 3.1.2: the existence of a radial gradient in the
morphology of And I's horizontal branch.  The sense of the gradient is such
that there are relatively more blue horizontal branch stars outside the 
core radius.  This gradient could reflect either an underlying
abundance gradient or a radial age gradient, but given the above discussion,
it would seem more likely that it should be interpreted as suggesting
that And I was more extended at its initial star forming
epoch, than when the bulk of its stars formed. 

In order to shed more light on this possibility, we have sought 
the existence of similar gradients in other 
dSph galaxies.  This task requires dSph studies that fulfill the following 
criteria: a) the available photometry must extend beyond the core radius;
b) the photometry must be reasonably precise at the magnitude of the 
horizontal branch; and c) the HB morphology must contain both blue and red
stars.  These criteria are satisfied by three studies: that of Carina by
Smecker-Hane {\it et al.\ }(1994), of Leo II by Demers \& 
Irwin (1993), and of Sculptor by \markcite{RL88}Light (1988).

Considering first Carina, the extensive study of 
Smecker-Hane {\it et al.\ }(1994) clearly separates the ``red clump'' 
of intermediate-age core helium
burning stars from the horizontal branch of old stars (both red and blue).
Smecker-Hane {\it et al.\ }(1994) find no difference
in the surface density distributions of these two classes of stars and further,
they indicate that there is no systematic trend with radius in the fraction
of the total core helium burning star population represented by the old 
HB stars.
These results apply out to a radius of $\sim$10$\arcmin$ beyond which their
sampling is incomplete.  Using the data of 
\markcite{IH95}Irwin \& Hatzidimitriou (1995, hereafter IH95), this limit
corresponds to $\sim$1.9 core radii on the major axis and $\sim$2.8 core
radii on the minor axis\footnote{We remind the reader that our use of the
term ``core radius'' differs from that of IH95, for whom the term refers to
the scale radius of the King (1966) model fit, rather than the radius at which
the surface density reaches half its central value.  The ratio of these two
length scales is a function of the central concentration of the best-fit
King model.  In all cases, we have corrected the IH95 ``core radius'' to our
convention by using the ratio appropriate for the central concentration listed
by IH95.}.  Hence, while the use by Smecker-Hane {\it et al.\ }(1994) of
circularly symmetric density profiles in this moderately flattened system
($\epsilon$ = 0.33, IH95) may have led to some loss of sensitivity to 
possible differences in the radial distributions of these two classes of stars,
it is apparent that Carina differs from And I in lacking a HB morphology
gradient.

For Leo II, we can analyze the data of Demers \& Irwin (1993) in much the same
way as was done for And I\@.  This dSph is circularly symmetric (IH95) and,
again using the results of IH95, the core radius is 
1.7$\arcmin$ $\pm$ 0.4$\arcmin$.  From a number of different techniques, the
center of this galaxy was found to be at (940, 1030) in the coordinate
system of Demers \& Irwin (1993). 
Then, defining red HB stars as those with
21.85 $<$ V $<$ 22.35 and 0.45 $\leq$ B--V $\leq$ 0.65, and blue HB stars
as those with the same V magnitude range, but with --0.2 $\leq$ B--V $\leq$ 
0.4 in the Demers \& Irwin (1993) cmd study, we find \mbox{{\it i} =
0.20 $\pm$ 0.02} for the 275 HB stars ({\it b} = 55, {\it r} = 220) inside
1.7$\arcmin$, and {\it i} = 0.22 $\pm$ 0.02 for the 396 HB stars ({\it b} = 
87, {\it r} = 309) outside this radius.  The T$_{2}$ statistic (defined above)
is 0.61 indicating
no statistically significant difference in the horizontal branch morphologies.
However, further investigation suggests that the HB morphology index 
for Leo II
is approximately constant out to r $\approx$ 3.0$\arcmin$, but beyond this
radius the fraction of blue HB stars increases.  For example, for 
r $\leq$ 3.0$\arcmin$, {\it i} = 0.20 $\pm$ 0.02 ({\it b} = 112, {\it r} = 454)
while for r $\geq$ 3.0$\arcmin$, {\it i} = 0.29 $\pm$ 0.04 ({\it b} = 30, 
{\it r} = 75).  Comparing these values yields T$_{2}$ = 2.02, or 
only a 2\% probability of
similar underlying HB morphologies.  This result does not change significantly
if the adopted center is changed by up to 20 pixels in either coordinate.
Thus, while the statistical significance
is not as high as it is for And I, there is a strong hint that the effect 
seen in And I is also present in Leo II\@.  Unfortunately, the HST data of
Mighell \& Rich (1996) do not reach far enough from the center of Leo II
to investigate whether or not the increased number of blue HB stars is
accompanied by an increase of older stars in the vicinity of the main sequence
turnoff.

Turning now to Sculptor, we note first that Da Costa (1984)
pointed out that in his cmd, which was for a small area well outside the core,
there were more blue HB stars than expected from scaling the cmd of 
\markcite{KD77}Kunkel \& Demers (1977), which applies to the central
regions of this dSph.  This result was followed up by Light (1988) who obtained
intermediate band Gunn system CCD photometry at a number of locations 
in Sculptor.
In his analysis, Light (1988) considered four circularly symmetric regions and
calculated for each the ratio of the number of stars in two distinct areas
in his cmds.  The first area contains the blue HB population (``blue'' stars)
while the
second contains contributions from both the red HB and from the red giant 
branch (``red'' stars).  This situation was made necessary by the fact that the
errors in his photometry did not permit a clean separation of the red HB
from the red giant branch population at similar magnitudes.  
Further, again because of photometric precision concerns, the magnitude
interval employed in defining these samples was overly generous, permitting
additional numbers of red giant branch stars to contaminate the samples.  The
inclusion of such stars will dilute the effect of any radial gradient that is
manifest only in the HB stars, as is the case in And I\@.  

Based on the data of IH95, the core radius of Sculptor is 
310$\arcsec$ $\pm$ 80$\arcsec$ on the major axis and 
210$\arcsec$ $\pm$ 60$\arcsec$ on the minor axis.  Consequently, Light's
region 1 (r $\leq$ 210$\arcsec$) is within the dSph's core radius for all
position angles.  Similarly, Light's regions 3 
(330$\arcsec$ $\leq$ r $\leq$ 480$\arcsec$) and 4 (r $\geq$ 480$\arcsec$)
are completely outside the core.  For region 1, there are 42 stars in the
blue star area in the cmd and 234 in the red star area leading to a
morphology index \mbox{{\it i} = 0.15 $\pm$ 0.02}.  
For regions 3 and 4 together,
blue HB stars are relatively more frequent with 
\mbox{{\it i} = 0.20 $\pm$ 0.03}
({\it b} = 45, {\it r} = 182).  The T$_{2}$ statistic is 1.36 yielding a
9\% probability that the samples are drawn from the same underlying 
distribution.  If we consider only the outermost region, then again 
the blue HB stars are more frequent;
\mbox{{\it i} = 0.28 $\pm$ 0.07} ({\it b} = 10, {\it r} = 26).  Comparing with 
region 1, the T$_{2}$ statistic is 1.91 for a 3\% probability that the samples
are from similar distributions.  Given the dilution effect of the inclusion
of red giant branch stars in these samples, we regard these results as
significant and conclude that, like And I and Leo II, Sculptor possesses 
a radial gradient
in its HB morphology.  Light (1988) reached the same conclusion.  Clearly
though, an extensive cmd study of Sculptor is required to place this conclusion
on a firmer footing.

Thus in 2 of the 3 galactic dSphs considered, there are indications of 
similar HB morphology gradients to that seen in And I\@.  Such a result should
not be too surprising given the overall similarity between the M31 dSphs and
the galactic dSph companions (e.g.\ Armandroff 1994).  Nevertheless, the 
existence
of a radial gradient in And I indicates that this dSph was more
extended at its earliest star formation epoch.  This in turn
argues rather forcefully that And I formed and evolved as an independent 
system,
rather than as debris from the disruption of a larger system (as is sometimes
argued for some, or indeed all, of the galactic dSphs, 
e.g. \markcite{LB95}Lynden-Bell \& Lynden-Bell 1995 and references therein).

\section{Summary}

In this paper we have presented results derived from WFPC2 images of 
the M31 dwarf spheroidal companion Andromeda I\@.  The analysis has revealed
for the first time the morphology of the horizontal branch in a dSph system
beyond those that are satellites of our Galaxy.  The And I HB morphology
is predominantly red, which combined with our metal abundance estimate 
of [Fe/H] = --1.45 $\pm$ 0.2 for this galaxy,
indicates that And I can be classified as a second parameter object lying in
the outer halo of M31.  This result then offers support for the hypothesis 
that the halos of M31 and the Galaxy formed in a similar manner.  The age of
the majority population in And I is estimated as $\sim$10 Gyr, though the
occurrence of blue HB stars and RR Lyrae variables in the cmd is an indication
of a minority population that is perhaps $\sim$3 Gyr or
more older.  And I is therefore again similar to the galactic dSphs in having
clear indications of an extended star formation history.  The WFPC2
observations extend far enough from the center of And I to permit the 
discovery of a radial gradient in the HB morphology: the blue HB stars and
RR Lyrae variables are relatively more common beyond And I's core.  This
result is interpreted as indicating that And I contracted between the epoch
of the initial star formation episode and the time
when the majority of the stars in the dSph formed.  Similar HB morphology
gradients were detected in two of three galactic dSphs studied.
We have also derived the line-of-sight distance between And I and M31 as 
0 $\pm$ 70 kpc and have verified the presence of an internal metal abundance
spread within this dSph galaxy.  The size of this metal abundance 
spread, however,
is somewhat smaller than might have been expected on the basis of And I's
luminosity.  It will be interesting to see how these And I results compare
with those for And II, a second M31 dSph companion which we will study with
HST in Cycle 6.

\acknowledgements
G. Da C. would like to dedicate this paper to the memory of his father, who
died while it was being written.  His interest and support over the years will
not be forgotten.  G. Da C. is grateful for travel support from the Australian
Government International Science \& Technology Program ``Space Science
with the Hubble Space Telescope'' and to Andrew Drake for discussions 
concerning the variable star periods.  This research was also supported in part
by NASA through grant number GO-05325 from the Space Telescope Science
Institute, which is operated by AURA, Inc., under NASA contract NAS 5-26555.

\end{document}